\documentclass[lettersize,journal]{IEEEtran}
\usepackage{amsmath,amsfonts}
\usepackage{algorithmic}
\usepackage{algorithm}
\usepackage{array}
\usepackage{textcomp}
\usepackage{stfloats}
\usepackage{url}
\usepackage{verbatim}
\usepackage{graphicx}
\usepackage{cite}
\usepackage{multirow}
\usepackage{amsmath} % Required for mathematical symbols and environments
\usepackage{graphicx}
\usepackage{subfigure}
\usepackage[table]{xcolor}  
\usepackage{xcolor}
\colorlet{Mycolor1}{green!10!orange}
\definecolor{mygreen}{RGB}{90, 200, 0}
\definecolor{myblue}{RGB}{90, 90, 250}
\definecolor{myyellow}{RGB}{190, 190, 0}
\definecolor{myred}{RGB}{200, 50, 0}
\usepackage{lipsum}
\usepackage[hidelinks]{hyperref}
\usepackage{booktabs}
\usepackage[english]{babel}
\usepackage{graphicx, type1cm, lettrine}
\usepackage[export]{adjustbox}
\usepackage{tikz}

\usepackage{lscape}
\usepackage{longtable}
\usepackage[table]{xcolor}
\usepackage{multicol,multirow}

\newcommand{\blue}[1]{\textcolor{myblue}{#1}}
\newcommand{\green}[1]{\textcolor{mygreen}{#1}}
\newcommand{\yellow}[1]{\textcolor{myyellow}{#1}}
\newcommand{\red}[1]{\textcolor{myred}{#1}}

\usepackage[figurename=Fig.,font=footnotesize,labelsep=none]{caption}
\setlipsum{%
  par-before = \begingroup\color{lightgray},
  par-after = \endgroup
}

\begin{document}
\nocite{*}

\title{Deep Interactive Segmentation of Medical Images: A Systematic Review and Taxonomy
}

\author{Zdravko Marinov$^*$, Paul F. Jäger$^*$, Jan Egger, Jens Kleesiek, Rainer Stiefelhagen

        % <-this % stops a space
\thanks{}% <-this % stops a space
\thanks{
This work has been submitted to the IEEE for possible publication. Copyright may be transferred without notice, after which this version may no longer be accessible.

%Manuscript received 21 November 2023. Date of current version 21 November 2023. 
The present contribution is supported by the Helmholtz Association under the joint research school “HIDSS4Health – Helmholtz Information and Data Science School for Health." \textit{($^*$Zdravko Marinov
and $^*$Paul F. Jäger are co-first authors.) (Corresponding author: Zdravko Marinov.)}

Zdravko Marinov and Rainer Stiefelhagen are with the Computer Vision for Human-Computer Interaction
Lab, Department of Informatics, Karlsruhe Institute of Technology, Adenauerring 10, 76131 Karlsruhe, Germany (e-mail: \url{zdravko.marinov@kit.edu}).

Jan Egger and Jens Kleesiek are with the Institute for Artificial Intelligence in Medicine
(IKIM), University Hospital Essen (AöR), Girardetstraße 2, 45131 Essen,
Germany.

Paul F. Jäger is with the German Cancer Research Center (DKFZ) Heidelberg, Interactive Machine Learning Group, Im Neuenheimer Feld 223, 69120 Heidelberg, Germany, and with the Helmholtz Imaging, DKFZ, Im Neuenheimer Feld 223, 69120 Heidelberg, Germany.
 }}

\markboth{PREPRINT VERSION JANUARY 2024}{PREPRINT VERSION JANUARY 2024}


\maketitle

\begin{abstract}
Interactive segmentation is a crucial research area in medical image analysis aiming to boost the efficiency of costly annotations by incorporating human feedback. This feedback takes the form of clicks, scribbles, or masks and allows for iterative refinement of the model output so as to efficiently guide the system towards the desired behavior. In recent years, deep learning-based approaches have propelled results to a new level causing a rapid growth in the field with 121 methods proposed in the medical imaging domain alone. 
In this review, we provide a structured overview of this emerging field featuring a comprehensive taxonomy, a systematic review of existing methods, and an in-depth analysis of current practices. Based on these contributions, we discuss the challenges and opportunities in the field. For instance, we find that there is a severe lack of comparison across methods which needs to be tackled by standardized baselines and benchmarks. Appendixes are \href{https://github.com/Zrrr1997/interactive_med_seg_review/blob/main/T_PAMI_Survey___Interactive_Segmentation___Appendix_Only.pdf}{here}.


\end{abstract}

\section{Introduction}
\lettrine[lines=2, findent=3pt, nindent=0pt]{D}{eep learning} segmentation methods revolutionized various application areas including autonomous driving \cite{kaymak2019brief}, product manufacturing \cite{tabernik2020segmentation}, and medical image analysis \cite{litjens2017survey}. For the latter, high-quality segmentation of anatomical structures and detection of abnormalities is an essential step towards automating diagnosis and treatment planning \cite{bakator2018deep}. However, the quality of these methods relies heavily on large-scale data sets for training featuring high-quality annotations. Especially in the medical imaging domain, this poses a major bottleneck, because annotations are time-consuming and require expert knowledge \cite{wang2018deepigeos}. For instance, labeling a volumetric Positron Emission Tomography/Computed Tomography (PET/CT) volume to identify tumor lesions can consume up to an hour of manual annotation for a single sample \cite{gatidis2023autopet}. 

Deep interactive segmentation addresses this trade-off between high-quality segmentation and laborious manual annotation. The idea is to boost annotation efficiency by incorporating human feedback into either the training or application process of segmentation methods. This feedback loop lets users iteratively correct or refine the model output, e.g. in the form of clicks, scribbles, or fine-grained voxel-masks, and thus efficiently guide the model towards the desired output.

The field of interactive segmentation traces back to active contour models \cite{kass1988snakes} and Graph Cut \cite{boykov2001interactive}, which primarily rely on low-level image features, such as pixel intensity changes, to differentiate foreground and background. However, these traditional methods face challenges when dealing with ambiguous boundaries and do not incorporate high-level semantics related to the object-of-interest \cite{xu2016deep,wang2018deepigeos,asad2022econet}. These challenges have been widely solved in recent years by interactive deep learning-based approaches, as first introduced by Xu et al. \cite{xu2016deep}. This paradigm shift has led to the successful application of interactive segmentation systems, for instance by reducing the annotation time of the aforementioned PET/CT volume to around three minutes \cite{hallitschke2023multimodal}. 

\begin{figure}[!t]
    \centering
    \includegraphics[width=\linewidth]{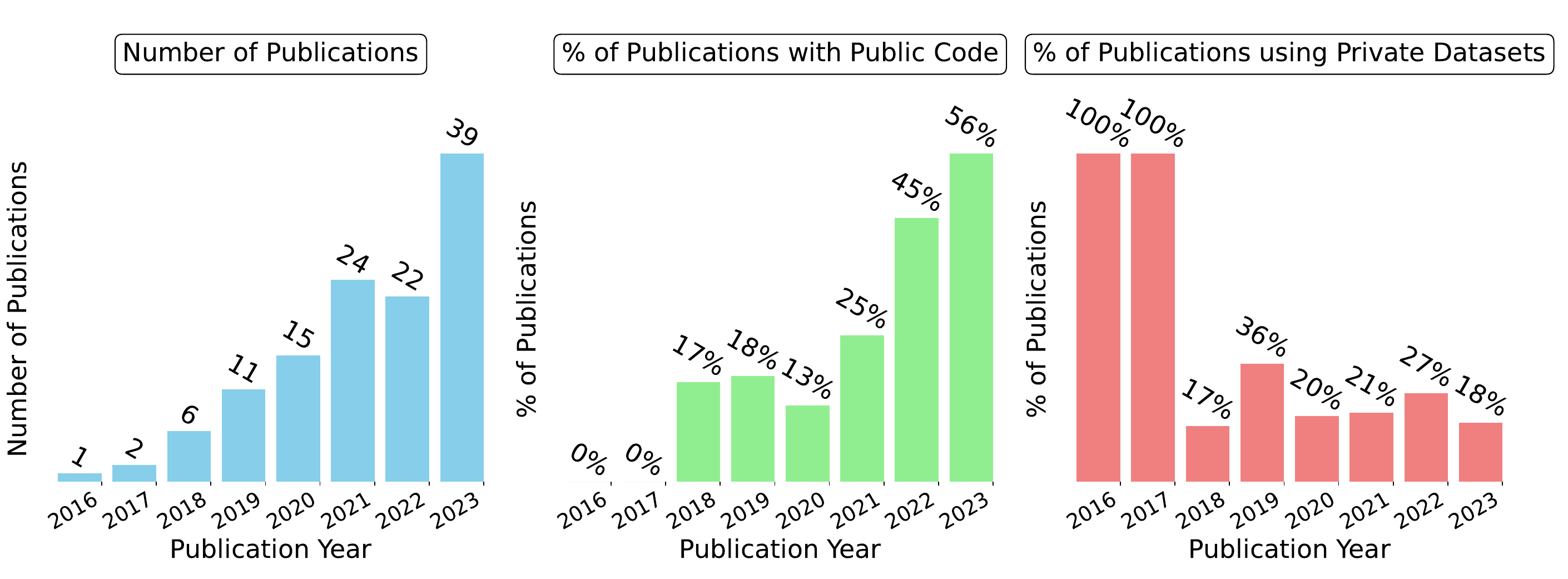}
    \caption{Tendencies in medical interactive segmentation in recent years.}
    \label{fig:num_pub}
\end{figure}

Several \label{rationale}reviews have been published in the field of interactive segmentation. However, previous reviews either focus on classical approaches rather than the more recent deep learning methods \cite{zhao2013overview,olabarriaga2001interaction,mcguinness2010comparative, 8215985}, or exclude approaches from the medical domain \cite{ramadan2020survey}. At the same time, no review exists for the field of deep learning-based interactive segmentation of medical images despite its rapid emergence with over 121 proposed methods in the last 8 years as seen in Fig. \ref{fig:num_pub}. The lack of a systematic overview in this field hampers scientific progress by generating redundancies and poses a challenge for users seeking the best-fitting method for their problem. 

We \label{objectives}address these shortcomings in this dedicated review by means of the following key contributions:

\begin{itemize}
    \item We introduce a comprehensive taxonomy for deep interactive segmentation allowing users to quickly comprehend the various approaches and select the best fitting method for their task.
    \item Based on this taxonomy, we provide a systematic review of 121 proposed methods in the medical domain.
    \item We perform an in-depth analysis of the current practices in the field including prevalent datasets, anatomies, and validation metrics, as well as the adequacy of baselines and the reproducibility of results.
    \item Based on this analysis, we provide a discussion of current challenges and opportunities in the field.
\end{itemize}

\section{Terminology}
Before we present our systematic review, we establish clear definitions for the fundamental terminology within the domain of interactive segmentation.
\subsection{Interactive Segmentation}
Interactive segmentation describes an iterative feedback loop, where user-provided corrections or refinements to the model's output inform subsequent iterations, leading to updated predictions. Depending on the method, user guidance is provided during training or application in the form of, e.g., clicks, scribbles, or other interactions. Importantly, initial labels provided to a model before training are excluded from this definition to differentiate interactive segmentation from related training paradigms such as weakly-supervised segmentation. 

\subsection{Guidance Signal} A guidance signal is a representation of the user interactions in a form in which the model can process it. This can be an explicit representation that involves transforming the user interaction into an additional structured input for the model to process and learn from, e.g., Gaussian heatmaps centered around user clicks. Additionally, guidance signals can also be implicit, where user interaction information is subtly integrated into the model's learning process without the provision of explicit structured input. For instance, this integration could involve modifying the loss function to incorporate the distance to clicks and assign greater weight to predictions in proximity to those clicks. Existing guidance signals for clicks, scribbles, and other interactions are given in the Appendixes.

\subsection{Training and Application} \label{sec:train_application} We use the terms \textit{training} and \textit{application} as the building blocks of our taxonomy tree. In the training stage, the model undergoes optimization, where its weights are updated using a predetermined loss function. The subsequent application stage involves deploying the trained model on unseen data, utilizing its refined parameters to address specific clinical tasks.

\subsection{Robot User} The concept of a robot user \cite{nickisch2010learning} involves creating a simulated model that mimics the behavior of a real human annotator. The robot user leverages ground-truth labels to simulate user interactions at plausible locations. For example, clicks can be sampled randomly from the ground-truth labels or generated at the center of the largest object. 
These simulated interactions are then converted into a guidance signal, which is fed back to the model. Robot users are used during training to simulate interactions for a large number of training samples as this is unfeasible for real human annotators at this scale. Additionally, robot users can also be used during application to evaluate trained models on unseen data without involving real human annotators.
Robot users can be categorized as non-iterative or iterative. Non-iterative users simulate all interactions simultaneously, integrate them into the model, and perform a single prediction. In contrast, iterative users simulate interactions in a loop. In this case, the model predicts, interactions are generated based on the errors of this prediction, and the model predicts again using all the previous interactions in an interaction-prediction loop \cite{mahadevanitis}. Here, an iteration denotes a single round of interaction and prediction with the model.

\iffalse
Robot users are usually used during training to simulate interactions for a large number of training samples but can also be used during application to quickly evaluate trained models without involving real human annotators.
\fi

\section{Scope and Study Collection Strategy}

\begin{figure}[t!]
    \centering
    \includegraphics[width=\linewidth]{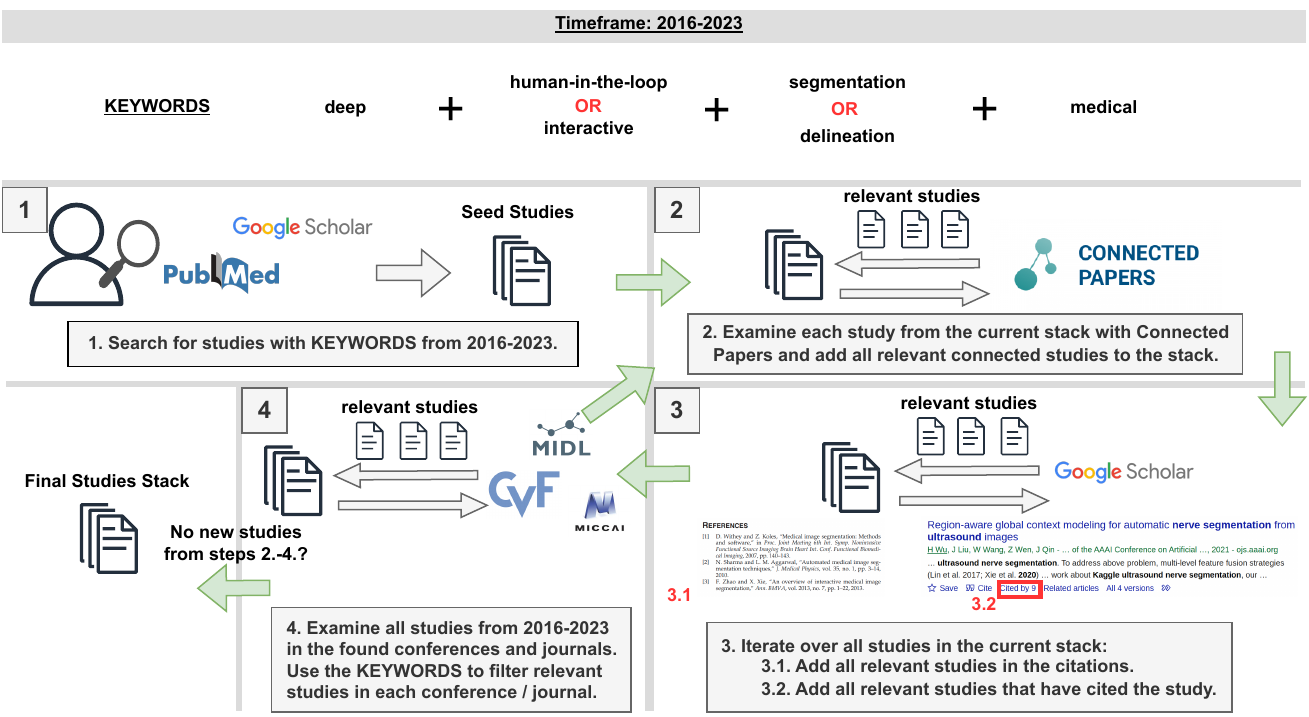}
    \caption{Search strategy in our systematic review for selecting relevant studies. The logos in steps 1 and 4 are illustrated only as examples for literature databases and venues respectively. A full list is given in the Appendixes.}
    \label{fig:scope}
\end{figure}

We conduct a systematic review of deep learning-based interactive segmentation models applied in medical scenarios. Our review, being inherently technical in nature, aims to rigorously categorize and analyze relevant literature. Recognizing the need for a comprehensive reporting framework, we integrate as many components from the Preferred Reporting Items for Systematic Reviews and Meta-Analyses (PRISMA) guidelines by Moher et al. \cite{moher2009preferred} as applicable to enhance the transparency and methodological clarity of our study. A detailed account of the adopted PRISMA components can be found in the PRISMA 2020 checklist in the Appendixes. We performed a literature search in several databases, including PubMed, Google Scholar, IEEE Xplore, SpringerLink, and arXiv, using specific keywords -- [interactive], [human-in-the-loop], [segmentation], [delineation], [medical], and [deep]. The search was carried out on 31 July 2023, and we limited the publication period to cover the years 2016--2023 since the first deep learning interactive method originated in 2016 \cite{xu2016deep}. We removed duplicates, including pre-prints followed by their peer-reviewed versions. Subsequently, we conducted an initial manual screening of titles and abstracts to ensure that the selected studies are relevant. After this initial screening, full texts were retrieved and reviewed for eligibility based on specific inclusion criteria: 1) studies with English full texts; 2) studies that have undergone peer-review or have pre-prints submitted to the arXiv database; and 3)
studies describing the application of interactive segmentation models for a human medical purpose. Consequently, certain exclusion criteria were applied to maintain the focus and quality of the review: 1) studies lacking English full texts; 2) 
studies that utilize non-deep learning models; 3) studies that utilize interactive models solely on natural images; and 4)
studies using medical images but not as the primary focus.

One reviewer assessed a study's eligibility through a three-stage process. Initially, we examine the title to decide if it focuses on deep medical interactive segmentation. If the title is ambiguous, we read the abstract for confirmation. In cases where the abstract remains unclear, we read the entire study.

\begin{table*}[!t]
\caption{\\ \textsc{Assignment of the reviewed methods to their corresponding taxonomy nodes. App: Application, Sim: Simulated, Iter: Iterative}} \label{tab:paper_taxonomy}
\centering
\scalebox{0.575}{
\rowcolors{2}{gray!10}{white}

\begin{tabular}{l|lm{3cm}llm{6cm}m{2.5cm}c}
\toprule
\textbf{Paper} & \textbf{Year} & \textbf{Interaction} & \textbf{Guidance Signal} & \textbf{Taxonomy Node} & \textbf{Target} & \textbf{Modality} & \textbf{Paper Link} \\ \hline
DeepCut \cite{rajchl2016deepcut} & 2016 & Bounding Box & GT Bounding Box & \green{App Sim Non-Iter Rule-based} & fetal brain, fetal lungs & MRI & \href{https://ieeexplore.ieee.org/stamp/stamp.jsp?arnumber=7739993\&casa_token=3aI8RBhEM-EAAAAA:jeQxPVd9qnr\_\_JWl8z8BPFS5QSvCqFLaCVIljjZXb2OT7ZriQQ4VeRKxJ0RG7N2ZcBEYEKuelw}{Link} \\

UI-Net \cite{amrehn2017ui} & 2017 &Scribbles& Error Skeletonization & \green{App Sim Iter Uniform Error Sampling} & liver lesions & CT & \href{https://diglib.eg.org/xmlui/bitstream/handle/10.2312/vcbm20171248/143-147.pdf?sequence=1}{Link} \\

Sun et al.\cite{sun2017point} & 2017 &Clicks& Location Prior & \green{App Sim Non-Iter Rule-based} & prostate & MRI & \href{https://link.springer.com/chapter/10.1007/978-3-319-67389-9_26}{Link} \\

Can et al.\cite{can2018learning} & 2018 &Scribbles& GT Scribbles & \green{App Sim Non-Iter Rule-based} & cardiac structures, prostate & MRI & \href{https://link.springer.com/chapter/10.1007/978-3-030-00889-5_27}{Link}  \\

DeepIGeoS \cite{wang2018deepigeos} & 2018 &Scribbles& Geodesic Maps & \green{App Sim Non-Iter Sampling-based} & placenta, brain tumors & MRI & \href{https://ieeexplore.ieee.org/stamp/stamp.jsp?arnumber=8370732\&tag=1}{Link} \\

BIFSeg \cite{wang2018interactive} & 2018 &Scribbles& GT Scribbles & \blue{Online Fine-tuning} & placenta, kidneys, fetal brain and lungs, brain tumors & MRI & \href{https://ieeexplore.ieee.org/stamp/stamp.jsp?arnumber=8270673}{Link} \\

InterCNN \cite{bredell2018iterative} & 2018 &Scribbles& Subset of GT Mask & \green{App Sim Iter Uniform Error Sampling} & prostate & MRI & \href{https://link.springer.com/chapter/10.1007/978-3-030-00919-9_42}{Link} \\

Dhara et al.\cite{dhara2019segmentation} & 2018 &Scribbles& GT Scribbles & \blue{Online Fine-tuning} & brain tumors & MRI & \href{https://link.springer.com/chapter/10.1007/978-3-030-11723-8_11}{Link} \\

Tang et al.\cite{tang2018semi} & 2018 &Bounding Boxes& Implicit & \green{App Sim Non-Iter Rule-based} & lung nodules, liver lesions, lymph nodes & CT & \href{https://link.springer.com/chapter/10.1007/978-3-030-00937-3_47}{Link} \\

Sakinis et al.\cite{sakinis2019interactive} & 2019 &Clicks& Gaussian Heatmaps & \green{App Sim Iter Distance Transform-based Error Sampling} & colon cancer, spleen, kidneys, gallbladder, esophagus, liver, stomach, blood vessels, pancreas, adrenal glands & CT & \href{https://arxiv.org/pdf/1903.08205.pdf}{Link} \\

Zhou et al.\cite{zhou2019interactive} & 2019 &Scribbles& GT Scribbles & \green{App Sim Iter Rule-based Custom Rules} & brain tumors & MRI & \href{https://link.springer.com/chapter/10.1007/978-3-030-32248-9_37}{Link} \\

Khan et al.\cite{khan2019extreme} & 2019 &Clicks& Chebychev Maps & \green{App Sim Non-Iter Rule-based} & heart, aorta, trachea, esophagus & CT & \href{https://link.springer.com/chapter/10.1007/978-3-030-32245-8_8}{Link} \\

DeepIGeoSv2 \cite{lei2019deepigeos} & 2019 &Scribbles& Euclidean Maps & \green{App Sim Non-Iter Sampling-based} & brain stem, parotid, optic nerve, optic chasm & CT & \href{https://link.springer.com/chapter/10.1007/978-3-030-33642-4_7}{Link} \\

iW-Net \cite{aresta2019iw} & 2019 &Clicks& Attraction Field Map & \green{App Sim Non-Iter Rule-based} & lung nodules & CT & \href{https://link.springer.com/content/pdf/10.1038/s41598-019-48004-8.pdf}{Link} \\

Roth et al.\cite{roth2019weakly} & 2019 &Clicks& Gaussian Heatmaps & \green{App Sim Non-Iter Rule-based} & liver, spleen, prostate, cardiac structures & CT, MRI, US & \href{https://link.springer.com/chapter/10.1007/978-3-030-33642-4_5}{Link} \\

Ceronne et al.\cite{cerrone2019end} & 2019 &Clicks& Disks & \green{App Sim Non-Iter Samping-based} & neuron cells & Microscopy & \href{https://openaccess.thecvf.com/content_CVPR_2019/papers/Cerrone_End-To-End_Learned_Random_Walker_for_Seeded_Image_Segmentation_CVPR_2019_paper.pdf}{Link} \\

Zheng et al.\cite{zheng2019deep} & 2019 &Scribbles& Implicit & \red{App Non-interactive Training Post-processing} & pancreas & CT & \href{https://dl.acm.org/doi/pdf/10.1145/3364836.3364885?casa_token=BKT10lGzbgkAAAAA:EETMPHosn--ig1SvdaNVEo7dJHMTgAklx4mMZ6T_ylq3Z0SpquJlXTaA1rPBh2QmBanShP_W7_3Z}{Link} \\

Chao et al.\cite{chao2019radiotherapy} & 2019 &Scribbles& Implicit & \blue{Online Fine-tuning} & esophageal cancer & PET/CT & \href{https://arxiv.org/abs/1904.03086}{Link} \\

Längkvist et al.\cite{langkvist2019interactive} & 2019 & Full Slice Annotation & Subset of GT Mask & \blue{Online Full Training} & lung structures & CT & \href{https://arxiv.org/pdf/1904.11701.pdf}{Link} \\

Wang et al.\cite{wang2019interactive} & 2019 &Contour Correction& Polygon Vertices & \green{App Sim Non-Iter Sampling-based} & liver & CT & \href{https://link.springer.com/chapter/10.1007/978-3-030-35817-4_2}{Link} \\

Boers et al.\cite{boers2020interactive} & 2020 &Scribbles& Implicit & \blue{Online Fine-tuning} & pancreas & CT & \href{https://iopscience.iop.org/article/10.1088/1361-6560/ab6f99/meta?casa_token=uE1mGjAhKWkAAAAA:5CHbDxhv2PKxaKi_4Pty48gsnBn7WkIvc3r0hQFFjNDpewuZvoCe7vzgrHK0SHu-bl68JXHFHA}{Link} \\

UGIR \cite{wang2020uncertainty} & 2020 &Scribbles& Geodesic Maps & \green{App Sim Non-Iter Sampling-based} & fetal brain & MRI & \href{https://link.springer.com/chapter/10.1007/978-3-030-59719-1_28}{Link} \\

IterMRL \cite{liao2020iteratively} & 2020 &Clicks& Geodesic Maps & \green{App Sim Iter Rule-based Error Center} & brain tumors, cardiac structures, prostate & MRI & \href{https://openaccess.thecvf.com/content_CVPR_2020/papers/Liao_Iteratively-Refined_Interactive_3D_Medical_Image_Segmentation_With_Multi-Agent_Reinforcement_Learning_CVPR_2020_paper.pdf}{Link} \\

Raju et al.\cite{raju2020user} & 2020 &Clicks& Gaussian Heatmaps & \green{App Sim Non-Iter Rule-based} & liver & CT & \href{https://link.springer.com/chapter/10.1007/978-3-030-59710-8_45}{Link} \\

BS-IRIS \cite{ma2020boundary} & 2020 &Clicks& Geodesic Maps & \green{App Sim Iter Uniform Error Sampling} & brain tumors, cardiac structures, prostate & MRI & \href{https://ieeexplore.ieee.org/stamp/stamp.jsp?arnumber=9311659}{Link} \\

NuClick \cite{koohbanani2020nuclick} & 2020 &Clicks + Scribbles& Disks + GT Skeletonization & \green{App Sim Non-Iter Sampling-based} & intestinal glands, cell nuclei, white blood cells & Microscopy & \href{https://www.sciencedirect.com/science/article/pii/S1361841520301353}{Link} \\

Kitrungrotsakul et al.\cite{kitrungrotsakul2020interactive} & 2020 &Scribbles& Error Skeletonization & \green{App Sim Iter Rule-based Error Skeletonization} & liver & CT & \href{https://arxiv.org/pdf/2006.15320.pdf}{Link} \\

IRIS \cite{pepe2020iris} & 2020 &Clicks& Implicit & \red{App Non-interactive Training Post-processing} & aorta & CTA & \href{https://www.spiedigitallibrary.org/conference-proceedings-of-spie/11317/113170R/IRIS--interactive-real-time-feedback-image-segmentation-with-deep/10.1117/12.2551354.short?SSO=1}{Link} \\

Hu et al.\cite{hu2020error} & 2020 &Clicks& Geodesic Maps & \green{App Sim Iter Uniform Error Sampling} & brain tumors & MRI, CT & \href{https://link.springer.com/chapter/10.1007/978-3-030-59861-7_2}{Link} \\

Tian et al.\cite{tian2020graph} & 2020 &Contour Correction& Polygon Vertices & \green{App Sim Iter Rule-based Worst Vertex Correction} & prostate, cardiac structures & MRI & \href{https://aapm.onlinelibrary.wiley.com/doi/abs/10.1002/mp.14327}{Link} \\

Chao et al. 2 \cite{chao2020interactive} & 2020 &Scribbles& Implicit & \green{App Sim Iter Rule-based Worst Slice Correction} & nasopharangeal and esophageal cancer & PET/CT & \href{https://arxiv.org/pdf/2012.06873v1.pdf}{Link} \\

Tang et al. 2 \cite{tang2020one} & 2020 &Clicks& Euclidean Maps + Disks & \green{App Sim Non-Iter Sampling-based} & lung nodules, liver lesions, lymph nodes & CT & \href{https://link.springer.com/chapter/10.1007/978-3-030-59719-1_56}{Link} \\

Jinbo et al.\cite{jinbo2020development} & 2020 &Scribbles& Gaussian Heatmaps & \green{App Sim Iter Rule-based Error Skeletonization} & liver & CT & \href{https://ieeexplore.ieee.org/stamp/stamp.jsp?arnumber=9291740}{Link} \\

Girum et al.\cite{girum2020fast} & 2020 &Clicks& Implicit & \green{App Sim Non-Iter Rule-based} & prostate, cardiac structures & CT, US & \href{https://link.springer.com/article/10.1007/s11548-020-02223-x}{Link} \\

Ho et al.\cite{ho2020deep} & 2020 &Scribbles& GT Scribbles & \yellow{Train Active Learning} & osteosarcoma & Microscopy & \href{https://link.springer.com/chapter/10.1007/978-3-030-59722-1_52}{Link} \\

Foo et al.\cite{foo2021interactive} & 2021 &Scribbles& Lines & \green{App Sim Iter Rule-based Error Center} & COVID19 lesions & CT & \href{https://arxiv.org/pdf/2110.00948.pdf}{Link} \\

Menon et al.\cite{menon2021interactive} & 2021 &Scribbles& Implicit & \yellow{Train Active Learning} & colorectal cancer, breast cancer & Microscopy & \href{https://link.springer.com/chapter/10.1007/978-3-031-02444-3_38}{Link} \\

MIDeepSeg \cite{luo2021mideepseg} & 2021 &Clicks& Exponential Geodesic Maps & \green{App Sim Non-Iter Rule-based} & placenta, spleen, kidney, prostate, fetal brain & CT, MRI, US & \href{https://www.sciencedirect.com/science/article/abs/pii/S1361841521001481}{Link} \\

Feng et al.\cite{feng2021interactive} & 2021 &Clicks& Disks & \green{App Sim Iter Rule-based Error Center} & liver, kidney, stomach, breast & CT & \href{https://ieeexplore.ieee.org/abstract/document/9358206}{Link} \\

Roth et al. 2 \cite{roth2021going} & 2021 &Clicks& Gaussian Heatmaps & \green{App Sim Non-Iter Rule-based} & spleen, liver, pancreas, kidneys, gallbladder & CT & \href{https://www.mdpi.com/2504-4990/3/2/26}{Link} \\

Sambaturu et al.\cite{sambaturu2021efficient} & 2021 &Scribbles& GT Scribbles & \blue{Online Fine-tuning} & cell nuclei, liver, liver lesions, heart, trachea, aorta, esophagus, brain tumors & Microscopy, MRI, CT & \href{https://link.springer.com/chapter/10.1007/978-3-030-87196-3_58}{Link} \\

Zhou et al. 2 \cite{zhou2021quality} & 2021 &Scribbles& GT Scribbles & \red{App Non-interactive Training Pre-saved Weak Labels} & lung, colon, kidney, kidney tumors & CT & \href{https://link.springer.com/chapter/10.1007/978-3-030-87196-3_52}{Link} \\

Williams et al.\cite{williams2021interactive} & 2021 &Contour Correction& B-splines & \red{App Non-interactive Training Post-processing} & levator hiatus & US & \href{https://www.springerprofessional.de/en/interactive-segmentation-via-deep-learning-and-b-spline-explicit/19687882}{Link} \\

WDTISeg \cite{li2021wdtiseg} & 2021 &Clicks& Euclidean Maps + Geodesic Maps & \green{App Sim Non-Iter Sampling-based} & breast cancer & US & \href{https://www.mdpi.com/2076-3417/11/14/6279}{Link} \\

Li et al.\cite{li2021interactive} & 2021 &Clicks& Gaussian Heatmaps & \green{App Sim Iter Uniform Error Sampling} & brain tumors, cardiac structures, spleen, liver & MRI, CT, CT/MRI & \href{https://link.springer.com/article/10.1631/FITEE.2200299}{Link} \\

Deng et al.\cite{deng20213d} & 2021 &Scribbles& Implicit & \green{App Sim Iter Uniform Error Sampling} & aortic system, brain tumors & CTA, MRI & \href{https://ieeexplore.ieee.org/document/9611087/}{Link} \\

Zhang et al.\cite{zhang2021interactive} & 2021 &Clicks& Implicit & \green{App Sim Non-Iter Rule-based} & kidney tumors, prostate & CT, MRI & \href{https://www.sciencedirect.com/science/article/pii/S093336572031263X or https://arxiv.org/pdf/1804.10481.pdf}{Link} \\

Zheng et al. 2 \cite{zheng2021continual} & 2021 &Scribbles& Implicit & \blue{Online Fine-tuning} & skin lesions & Dermoscopy & \href{https://ojs.aaai.org/index.php/AAAI/article/view/16752}{Link} \\

DINs \cite{zhang2021dins} & 2021 &Clicks& Gaussian Heatmaps & \green{App Sim Iter Rule-based Error Center} & neurofibromatosis type I & MRI & \href{https://ieeexplore.ieee.org/abstract/document/9449950/}{Link} \\

Tian et al. 2 \cite{tian2021interactive} & 2021 &Contour Correction& Polygon Vertices & \green{App Sim Iter Rule-based Worst Vertex Correction} & prostate & MRI & \href{https://www.sciencedirect.com/science/article/abs/pii/S0925231221001053}{Link} \\

Jiang et al.\cite{jiang2021residual} & 2021 &Clicks& Gaussian Heatmaps & \green{App Sim Non-Iter Sampling-based} & skin lesions & Dermoscopy & \href{https://link.springer.com/article/10.1186/s13326-021-00255-z}{Link} \\

Bai et al.\cite{bai2021progressive} & 2021 &Clicks& Gaussian Heatmaps & \green{App Sim Iter Prediction-based Error Sampling} & cardiac structures, prostate, kidney, spleen, lung cancer, colon cancer, kidney cancer & CT, MRI & \href{https://www.spiedigitallibrary.org/conference-proceedings-of-spie/11596/115962Q/Progressive-medical-image-annotation-with-convolutional-neural-network-based-interactive/10.1117/12.2582241.short}{Link} \\

Deepscribble \cite{cho2021deepscribble} & 2021 &Scribbles& Euclidean Maps & \green{App Sim Iter Rule-based Error Skeletonization} & liver tumors & Microscopy & \href{https://ieeexplore.ieee.org/abstract/document/9434105}{Link} \\

Attention-RefNet \cite{kitrungrotsakul2021attention} & 2021 &Scribbles& Geodesic Maps & \green{App Sim Iter Rule-based Error Skeletonization} & COVID19 lesions & CT & \href{https://ieeexplore.ieee.org/stamp/stamp.jsp?arnumber=9440763\&casa_token=8XXbVKFR0H4AAAAA:5L2i_VnSn9tIYIw20gUmX9Jit_RsIMuPVwJZcRDSmvWAMQ6FRrX7ybpjou249ni_PpcNVCCx}{Link} \\

Daulatabad et al.\cite{daulatabad2021integrating} & 2021 &Clicks& Disks & \green{App Sim Non-Iter Sampling-based} & thyroid nodules & US & \href{https://ieeexplore.ieee.org/abstract/document/9629959?casa_token=ITSjjs-PloAAAAAA:gEo0ksYw8P75cDPINO0w4_p2yuqpO6trt4j7k-ZpiqbhQQPwjG_mTCtIgz8wBkCf998B2bOg}{Link} \\

Manh et al.\cite{manh2021interactive} & 2021 &Clicks& Implicit & \red{App Non-interactive Training Post-processing} & Z-line & Colonoscopy & \href{https://ieeexplore.ieee.org/abstract/document/9642141/}{Link} \\

Trimpl et al.\cite{trimpl2021interactive} & 2021 & Full Slice Annotation & Subset of GT Mask & \green{App Sim Non-Iter Rule-based} & heart, esophagus, lungs, spinal cord, lung tumors, pancreas, pancreas tumors, liver, liver tumors, neck, submandibular gland, parotid, brain, brain tumors, spleen & CT & \href{https://aapm.onlinelibrary.wiley.com/doi/full/10.1002/mp.14852}{Link} \\

PiPo-Net \cite{fang2021pipo} & 2021 &Contour Correction& Polygon Vertices & \red{App Non-interactive Training Post-processing} & breast cancer & Microscopy & \href{https://ieeexplore.ieee.org/abstract/document/9636146?casa_token=VRiHMkyG0rEAAAAA:H6VH4hRld6_VrNx4gkWQZZpfy_zqEhK7yaKcq2PcEMYTejZAjpInGGHxDiMxFhORmUl1RwO9}{Link} \\

Jahanifar et al.\cite{jahanifar2021robust} & 2021 &Scribbles& GT Skeletonization & \green{App Sim Non-Iter Rule-based} & breast cancer & Microscopy & \href{https://openaccess.thecvf.com/content/ICCV2021W/CDPath/html/Jahanifar_Robust_Interactive_Semantic_Segmentation_of_Pathology_Images_With_Minimal_User_ICCVW_2021_paper.html}{Link} \\

Sun et al. 2 \cite{sun2022efficient} & 2022 &Contour Correction& B-splines & \red{App Non-interactive Training Post-processing} & prostate, nasoprahangeal cancer & MRI & \href{https://pubmed.ncbi.nlm.nih.gov/36433795/}{Link} \\

Shahedi et al.\cite{shahedi2022effect} & 2022 &Clicks& Disks & \green{App Sim Non-Iter Sampling-based} & prostate & CT & \href{https://aapm.onlinelibrary.wiley.com/doi/abs/10.1002/mp.15404?casa_token=pj2QCMFngH4AAAAA:P9JOfIHD_atCtB5r7iTFLeEWGJavzx2DOg0k9SdVGedD-5ZDkGVKM9o6V1FaXdid9V5Q538vyOC4tpY}{Link} \\

Atzeni et al.\cite{atzeni2022deep} & 2022 &Scribbles& Implicit & \yellow{Train Active Learning} & brain structures & MRI, Microscopy & \href{https://www.sciencedirect.com/science/article/pii/S1361841522001967}{Link} \\

Bi et al.\cite{bi2022hyper} & 2022 &Clicks& Euclidean Maps & \green{App Sim Non-Iter Sampling-based} & skin lesions & Dermoscopy & \href{https://www.sciencedirect.com/science/article/abs/pii/S1361841521003790}{Link} \\

iSegFormer \cite{liu2022isegformer} & 2022 &Clicks& Disks & \green{App Sim Iter Rule-based Error Center} & knee cartilage & MRI & \href{https://link.springer.com/chapter/10.1007/978-3-031-16443-9_45}{Link} \\

ECONet \cite{asad2022econet} & 2022 &Scribbles& GT Scribbles & \blue{Online Full Training} & COVID19 lesions & CT & \href{https://2022.midl.io/papers/i3}{Link} \\

i3Deep \cite{gotkowski2022i3deep} & 2022 &Scribbles& Subset of GT Mask & \green{App Sim Non-Iter Sampling-based} & brain tumors, pancreas, COVID19 lesions & CT, MRI & \href{https://2022.midl.io/papers/d_l_17}{Link} \\

DeepEdit \cite{diaz2022deepedit} & 2022 &Clicks& Gaussian Heatmaps & \green{App Sim Iter Prediction-based Error Sampling} & prostate, prostate tumors & MRI, CT, CT/MRI  & \href{https://link.springer.com/chapter/10.1007/978-3-031-17027-0_2}{Link} \\

Liu et al.\cite{liu2022transforming} & 2022 &Clicks& Disks & \green{App Sim Iter Rule-based Error Center} & lung, colon, pancreas & CT & \href{https://link.springer.com/chapter/10.1007/978-3-031-16440-8_67}{Link} \\

Shi et al.\cite{shi2022hybrid} & 2022 &Scribbles& Error Skeletonization & \green{App Sim Non-Iter Sampling-based} & colon cancer, lung cancer, kidney tumors, kidney & CT & \href{https://link.springer.com/chapter/10.1007/978-3-031-16440-8_64}{Link} \\

AnatomySketch \cite{zhuang2022anatomysketch} & 2022 &Scribbles + Contour Correction& GT Scribbles + B-Splines & \yellow{Train Active Learning} & liver cancer, lung lobes, intervertebral disc & MRI, CT & \href{https://link.springer.com/article/10.1007/s10278-022-00660-5}{Link} \\

Galisot et al.\cite{galisot2022visual} & 2022 &Bounding Boxes& Implicit & \green{App Sim Non-Iter Rule-based} & brain structures & MRI & \href{https://www.sciencedirect.com/science/article/pii/S266682702200024X}{Link} \\

Lin et al.\cite{lin2022multi} & 2022 &Scribbles + Clicks + Bounding Boxes& Gaussian Heatmaps + BBox Mask & \green{App Sim Iter Rule-based Custom Rules} & COVID19 lesions, brain tumors, brachial plexus, polyps, skin lesions & X-Ray, CT, MRI, US, Colonoscopy & \href{https://dl.acm.org/doi/abs/10.1145/3503161.3548096}{Link} \\

Pirabaharan et al.\cite{pirabaharan2022interactive} & 2022 &Clicks& Gaussian Heatmaps & \green{App Sim Non-Iter Sampling-based} & spleen, colon cancer & CT & \href{https://ieeexplore.ieee.org/document/9871361}{Link} \\

Mikhailov et al.\cite{mikhailov2022deep} & 2022 &Clicks& Disks & \green{App Sim Iter Uniform Error Sampling} & uterus, bladder, uterine cavity, female pelvis tumors & MRI & \href{https://link.springer.com/chapter/10.1007/978-3-031-17721-7_11}{Link} \\

Pirabaharan et al. 2 \cite{pirabaharan2022improving} & 2022 &Clicks& Gaussian Heatmaps & \green{App Sim Non-Iter Sampling-based} & spleen, colon cancer & CT & \href{https://ieeexplore.ieee.org/abstract/document/9999079?casa_token=cCmuxB2ObAAAAAAA:K3jNxizmihJXd5WQQnJXT0fH6fcjk5-GZYMSp4MxEZ9OOixn8O2SZiXwUcTNUsC_ssUm1qPr}{Link} \\

Chen et al.\cite{chen2022balancing} & 2022 &Clicks& Gaussian Heatmaps + Euclidean Maps & \green{App Sim Non-Iter Rule-based} & breast lesions & US & \href{https://www.sciencedirect.com/science/article/abs/pii/S1746809422002452}{Link} \\

Deep SED-Net \cite{liang2022deep} & 2022 &Scribbles& GT Scribbles & \yellow{Train Active Learning} & testicular cells & Microscopy & \href{https://onlinelibrary.wiley.com/doi/abs/10.1002/cyto.a.24556?casa_token=TP_TRBtnr0sAAAAA:42gHNTdNC4MEl4X-9XJlBlZPPbKQpb5qgh4JVzt8SjW3s9JYjlxZbQ6ast1cFUeoVtU6hlxs4PkvXaQ}{Link} \\

Ju et al.\cite{ju2022all} & 2022 &Clicks& Implicit & \green{App Sim Non-Iter Sampling-based} & liver, kidneys, spleen & CT & \href{https://www.mdpi.com/2076-3417/12/3/1328}{Link} \\

Ma et al.\cite{ma2022rapid} & 2022 &Scribbles& Implicit & \yellow{Train Active Learning} & liver, spleen & CT, MRI & \href{https://ieeexplore.ieee.org/document/9761467}{Link} \\

Bai et al. 2 \cite{bai2022proof} & 2022 &Clicks& Gaussian Heatmaps & \green{App Sim Iter Prediction-based Error Sampling} & nasopharangeal cancer & CT & \href{https://pubs.rsna.org/doi/abs/10.1148/ryai.210214}{Link} \\

Zhou et al. 3 \cite{zhou2023volumetric} & 2023 &Scribbles& Geodesic Maps & \red{App Non-interactive Training Pre-saved Weak Labels} & lung, colon, kidney, kidney tumors & CT, Colonoscopy & \href{https://www.sciencedirect.com/science/article/pii/S1361841522002316}{Link} \\

Hallitschke et al.\cite{hallitschke2023multimodal} & 2023 &Scribbles& Geodesic Maps & \green{App Sim Non-Iter Sampling-based} & lung cancer & PET/CT & \href{https://arxiv.org/pdf/2301.09914.pdf}{Link} \\

Liu et al. 2 \cite{liu2023exploring} & 2023 &Clicks + Scribbles& Disks + GT Scribbles & \green{App Sim Iter Rule-based Error Center} & esophagus, cardiac structures, pancreas, celiac trunk, spleen, gallbladder, stomach, kidney, liver, aorta, ribs, femoral cartilage & CT & \href{https://arxiv.org/abs/2303.06493}{Link} \\

Bruzadin et al.\cite{bruzadin2023learning} & 2023 &Scribbles& Implicit & \green{App Sim Non-Iter Rule-based} & COVID19 lesions & CT & \href{https://www.sciencedirect.com/science/article/pii/S0925231222015041}{Link} \\

Asad et al.\cite{asad2023adaptive} & 2023 &Scribbles& GT Scribbles & \blue{Online Full Training} & COVID19 lesions & CT & \href{https://arxiv.org/pdf/2303.13696v1.pdf}{Link} \\

Shahin et al.\cite{shahin2023sparse} & 2023 &Scribbles& Gaussian Heatmaps & \green{App Sim Non-Iter Rule-based} & cardiac structures & US & \href{https://arxiv.org/abs/2303.11041}{Link} \\

Zhuang et al.\cite{zhuang2023efficient} & 2023 &Contour Correction& Polygon Vertices & \yellow{Train Active Learning} & liver, spleen, kidneys & CT & \href{https://link.springer.com/article/10.1007/s11548-022-02730-z}{Link} \\

Ho et al. 2 \cite{ho2023deep} & 2023 &Scribbles& GT Scribbles & \yellow{Train Active Learning} & ovarian cancer & Microscopy & \href{https://www.sciencedirect.com/science/article/pii/S2153353922007544}{Link} \\

Wei et al.\cite{wei2023towards} & 2023 &Scribbles& GT Scribbles & \green{App Sim Iter Rule-based Worst Slice Correction} & head-and-neck cancer & PET/CT/MRI & \href{https://www.sciencedirect.com/science/article/pii/S2405631622001063\#s0105}{Link} \\

Zhuang et al. 2 \cite{zhuang2023annotation} & 2023 &Scribbles& Exponential Geodesic Maps & \yellow{Train Active Learning} & brain tumors, liver tumors & CT, MRI & \href{https://link.springer.com/article/10.1007/s11548-023-02931-0}{Link} \\

GtG \cite{marinov2023guiding} & 2023 &Clicks& Gaussian Heatmaps & \green{Sim Iter Distance Transform-based Sampling} & lung lesions, lymphoma, melanoma, spleen & CT, PET/CT & \href{https://arxiv.org/abs/2303.06942}{Link} \\

Qu et al.\cite{qu2023abdomenatlas} & 2023 &Scribbles& Subset of GT Mask & \yellow{Train Active Learning} & spleen, liver, kidneys, stomach, gallbladder, pancreas, aorta, cardiac structures & CT & \href{https://arxiv.org/abs/2305.09666}{Link} \\ \bottomrule

\end{tabular}}
\end{table*}

%\clearpage
\begin{table*}[!t]
\caption{\\ \textsc{Assignment of SAM-based \cite{kirillov2023segment} methods to their taxonomy nodes. App: Application, Sim: Simulated, Iter: Iterative} \label{tab:paper_taxonomy_sam}}
\centering
\scalebox{0.6}{
\rowcolors{2}{gray!10}{white}

\begin{tabular}{l|lm{3cm}m{2.5cm}lm{7cm}m{4.5cm}c}
\toprule
\textbf{Paper} & \textbf{Year} & \textbf{Interaction} & \textbf{Guidance Signal} & \textbf{Taxonomy Node} & \textbf{Target} & \textbf{Modality} & \textbf{Paper Link} \\ \hline
Mazurowski et al. \cite{mazurowski2023segment} & 2023 & Clicks + Bounding Boxes & Positional Encoding & \green{App Sim Iter Uniform Error Sampling} & gray matter, spinal cord, heart, prostate, brain tumors, breasts, female genital tract, chest, ilium, femur, kidney, muscle, nerves, ovarian tumors, colon cancer, vessels, spleen, liver, bladder, lungs, lung cancer, melanoma, lymphoma & CT, MRI, US, X-Ray, PET/CT &  \href{https://arxiv.org/pdf/2304.10517.pdf}{Link} \\
Deng et al. \cite{deng2023segment} & 2023 & Clicks + Bounding Boxes & Positional Encoding & \green{App Sim Iter Uniform Error Sampling} & cell nuclei, skin cancer & Microscopy &  \href{https://arxiv.org/abs/2304.04155}{Link} \\

SAM vs. BET \cite{mohapatra2023sam} & 2023 & Clicks + Bounding Boxes & Positional Encoding & \green{App Sim Iter Uniform Error Sampling} & brain & MRI & \href{https://arxiv.org/abs/2304.04738}{Link} \\

Putz et al. \cite{putz2023segment} & 2023 & Clicks + Bounding Boxes & Positional Encoding & \green{App Sim Iter Uniform Error Sampling} & brain tumors & MRI & \href{https://arxiv.org/pdf/2304.07875.pdf}{Link} \\
Hu et al. \cite{hu2023sam} & 2023 & Clicks + Bounding Boxes & Positional Encoding & \green{App Sim Iter Uniform Error Sampling} & liver tumors & CT & \href{https://arxiv.org/pdf/2304.08506.pdf}{Link} \\
SAM-Adapter \cite{chen2023sam} & 2023 & Clicks + Bounding Boxes & Positional Encoding & \green{App Sim Iter Uniform Error Sampling} & polyps & Colonoscopy & \href{https://arxiv.org/pdf/2304.09148.pdf}{Link} \\
Medical SAM Adapter \cite{wu2023medical} & 2023 & Clicks &Positional Encoding & \green{App Sim Iter Uniform Error Sampling} & spleen, kidneys, gallbladder, esophagus, liver, stomach, aorta, cardiac structures, pancreas, adrenal glands, duodenum, bladder, optic cup, brain tumors, thyroid nodules & CT, MRI, US, Fundus, Dermoscopy & \href{https://arxiv.org/abs/2304.12620}{Link} \\

Ophthalmology SAM \cite{qiu2023learnable} & 2023 & Clicks + Bounding Boxes & Positional Encoding & \green{App Sim Iter Uniform Error Sampling} & blood vessels, retinal lesions & Fundus & \href{https://arxiv.org/pdf/2304.13425.pdf}{Link} \\

He et al. \cite{he2023accuracy} & 2023 & Clicks + Bounding Boxes & Positional Encoding & \green{App Sim Iter Uniform Error Sampling} & heart, brain, breasts, lung, bowel, pancreas, prostate, skin, heart, liver, brain, chest & MRI, US, CT, Colonoscopy, Dermoscopy, X-Ray & \href{https://arxiv.org/pdf/2304.12620.pdf}{Link} \\

Shi et al. \cite{shi2023generalist} & 2023 & Clicks + Bounding Boxes & Positional Encoding & \green{App Sim Iter Uniform Error Sampling} & skin, eyes, chest, colon, retina, abdominal organs & Dermoscopy, Fundus, CT, MRI, Colonoscopy, X-Ray, OCT & \href{https://arxiv.org/pdf/2304.12637.pdf}{Link} \\

GazeSAM \cite{wang2023gazesam} & 2023 & Eye Gaze & Positional Encoding & \green{App Sim Iter Uniform Error Sampling} & not specified & not specified & \href{https://arxiv.org/pdf/2304.13844.pdf}{Link} \\
SkinSAM \cite{hu2023skinsam} & 2023 & Bounding Boxes& Positional Encoding & \green{App Sim Iter Uniform Error Sampling} & skin lesions & Dermoscopy  & \href{https://arxiv.org/pdf/2304.13973.pdf}{Link} \\

Wang et al. \cite{wang2023sam} & 2023 & Clicks + Bounding Boxes & Positional Encoding & \green{App Sim Iter Uniform Error Sampling} & surgery instruments & Colonoscopy & \href{https://arxiv.org/pdf/2304.14674.pdf}{Link} \\
Cheng et al.. \cite{cheng2023sam} & 2023 & Clicks + Bounding Boxes & Positional Encoding & \green{App Sim Iter Uniform Error Sampling} & breasts, polyps, foot ulcers, COVID19 lesions, hippocampus, thyroid nodules, thyroid gland, liver tumors & US, Colonoscopy, MRI, CT & \href{https://arxiv.org/pdf/2305.00035.pdf}{Link} \\
Mattjie et al. \cite{mattjie2023exploring} & 2023 & Clicks + Bounding Boxes & Positional Encoding & \green{App Sim Iter Uniform Error Sampling} & skin lesions, lungs, femur, illium, polyps, breasts & X-Ray, US, Colonoscopy, Dermoscopy & \href{https://arxiv.org/pdf/2305.00109.pdf}{Link} \\
Polyp-SAM \cite{li2023polyp} & 2023 & Bounding Boxes& Positional Encoding & \green{App Sim Iter Uniform Error Sampling} & polyps & Colonoscopy & \href{https://arxiv.org/pdf/2305.00293.pdf}{Link} \\
PromptUNet \cite{wu2023promptunet} & 2023 & Clicks + Bounding Boxes + Scribbles& Positional Encoding + Subset of GT Mask & \green{App Sim Iter Uniform Error Sampling} & spleen, kidneys, gallbladder, esophagus, liver, stomach, aorta, cardiac structures, pancreas, adrenal glands, duodenum, bladder, optic cup, brain tumors, thyroid nodules & CT, MRI, US, Fundus, Dermoscopy & \href{https://arxiv.org/abs/2305.10300}{Link} \\
BreastSAM \cite{hu2023breastsam} & 2023 & Clicks + Bounding Boxes & Positional Encoding & \green{App Sim Iter Uniform Error Sampling} & breast cancer & US  & \href{https://arxiv.org/abs/2305.12447}{Link} \\
IAMSAM \cite{lee2023iamsam} & 2023 & Clicks + Bounding Boxes & Positional Encoding & \green{App Sim Iter Uniform Error Sampling} & breast cancer, colon, brain, prostate cancer,  & Microscopy & \href{https://www.biorxiv.org/content/10.1101/2023.05.25.542052.abstract}{Link} \\
DeSAM \cite{gao2023desam} & 2023 & Clicks + Bounding Boxes & Positional Encoding & \green{App Sim Iter Uniform Error Sampling} & prostate & MRI & \href{https://arxiv.org/abs/2306.00499}{Link} \\
Shen et al. \cite{shen2023temporally} & 2023 & Clicks + Bounding Boxes & Positional Encoding & \green{App Sim Iter Uniform Error Sampling} & brain tumors & MRI & \href{https://arxiv.org/pdf/2306.08958.pdf}{Link} \\
Ning et al. \cite{ning2023potential} & 2023 & Clicks + Bounding Boxes & Positional Encoding & \green{App Sim Iter Uniform Error Sampling} & heart, thyroid, carotid artery & US & \href{https://www.jstage.jst.go.jp/article/bst/advpub/0/advpub_2023.01119/_pdf}{Link} \\
Zhang et al. \cite{zhang2023segment} & 2023 & Clicks + Bounding Boxes & Positional Encoding & \green{App Sim Iter Uniform Error Sampling} & prostate, bladder, femoral heads, rectum, lungs, heart, spinal cord, esophagus, liver, stomach, kidneys, large and small bowels, brain, parotids, mandible, cochleas & CT & \href{https://arxiv.org/pdf/2306.11730.pdf}{Link} \\
MedLSAM \cite{lei2023medlsam} & 2023 & Clicks + Bounding Boxes & Positional Encoding & \green{App Sim Iter Uniform Error Sampling} & liver, spleen, kidneys, stomach, gallbladder, esophagus, pancreas, duodenum, colon, intestines, adrenal gland, rectum, bladder, femur heads, brain stem, eyes, lens, optic nerve, optic chiasm, brain structures, pituitary gland, parotids, cochlea, spinal chord, mandibles & CT & \href{https://arxiv.org/pdf/2306.14752.pdf}{Link} \\
SAM-U \cite{deng2023sam} & 2023 & Bounding Boxes& Positional Encoding & \green{App Sim Iter Uniform Error Sampling} & optic disc, optic cup & Fundus & \href{https://arxiv.org/pdf/2307.04973.pdf}{Link} \\ 
3DSAM-adapter \cite{gong20233dsam} & 2023 & Clicks &Positional Encoding & \green{App Sim Iter Uniform Error Sampling} & liver tumors, kidney tumors, pancreas tumors, colon cancer & CT & \href{https://arxiv.org/abs/2306.13465}{Link} \\

Huang and Yang et al. \cite{huang2023segment} & 2023 & Clicks + Bounding Boxes & Positional Encoding & \green{App Sim Iter Uniform Error Sampling} & See [\citenum{huang2023segment}, Table 2] for a comprehensive list of all 68 targets & CT, MRI, Colonoscopy, US, Fundus, Microscopy, Colonoscopy, X-Ray & \href{https://arxiv.org/pdf/2304.14660.pdf}{Link} \\
MedSAM \cite{ma2023segment} & 2023 & Clicks + Bounding Boxes & Positional Encoding & \green{App Sim Iter Uniform Error Sampling} & See Tables 1-4 in MedSAM's supplementary material (\href{https://github.com/bowang-lab/MedSAM/blob/main/assets/MedSAM_supp.pdf}{link}) for a comprehensive list of all $>$60 targets & CT, MRI, US, X-Ray, PET/CT, Microscopy, OCT, Colonoscopy, Fundus &  \href{https://arxiv.org/abs/2304.12306}{Link} \\

SAM.MD \cite{roy2023sam} & 2023 & Clicks + Bounding Boxes & Positional Encoding & \green{App Sim Iter Uniform Error Sampling} & spleen, kidneys, gallbladder, esophagus, liver, stomach, aorta, postcava, pancreas, adrenal glands, duodenum, bladder & CT & \href{https://arxiv.org/abs/2304.05396}{Link} \\
\bottomrule

\end{tabular}}
\end{table*}

%\clearpage

This search produced our initial \textit{seed studies} stack as illustrated in Fig. \ref{fig:scope}. In addition to adhering to the PRISMA guidelines, we implemented three supplementary steps in our search strategy to maximize the retrieval of relevant studies and formed an iterative loop utilizing these steps. These steps are depicted as steps 2, 3, and 4 in Fig. \ref{fig:scope}. In step 2, we incorporated the Connected Papers tool\footnote{\url{https://www.connectedpapers.com/}} to enhance our search process. This tool was applied to each of the already included studies from the \textit{seed studies}, and we systematically screened all the suggested studies recommended by the tool, ensuring they meet our predefined inclusion and exclusion criteria. In step 3, we manually inspected all the citations of each study in the \textit{seed studies} and all of the studies that have cited this study using the "Cited by" function in Google Scholar. In step 4, we formed a list of all the peer-reviewed venues, which is given in the Appendixes, and manually screened all of the publications from each venue in the timeframe 2016--2023 with our pre-defined keywords and added the relevant publications in our \textit{seed studies}. We repeated steps 2, 3, and 4 and accumulated all relevant studies in our \textit{seed studies} stack until no new relevant studies were found. Our search strategy found a total of 121 relevant publications. 

After collecting all studies, one reviewer manually extracted from each study the following data items: 1) used imaging modalities; 2) used datasets along with provided links, if available; 3) prior interactive methods the study has compared to; 4) employed evaluation metrics; 5) type of interaction, e.g., clicks; 6) target structures for segmentation; 7) and, if applicable, a link to publicly available code. We cataloged all 121 reviewed studies and their data items in Tables \ref{tab:paper_taxonomy} and \ref{tab:paper_taxonomy_sam}. This facilitates efficient navigation for future researchers seeking relevant interactive methods related to their own work.

It is important to note that during our search we exclude "classical approaches", which do not utilize deep learning. Some examples include methods based on: 1) Graph Cut \cite{boykov2001interactive, rother2004grabcut}; 2) dense Conditional Random Fields (CRFs) \cite{krahenbuhl2011efficient}; 3) active contours \cite{kass1988snakes}; and 4) level sets \cite{lefohn2003interactive}.  While non-deep learning interactive frameworks such as ilastik \cite{sommer2011ilastik} and ITK-Snap \cite{yushkevich2016itk} have demonstrated success in clinical workflows, we maintain a focus on deep learning-based methods to align with the review's scope and the growing prevalence of interactive deep learning models in the medical domain.

\section{Taxonomy}

\begin{figure*}[t]
    \centering
    \hspace*{-0.6cm}
    \includegraphics[width=1.0\linewidth]{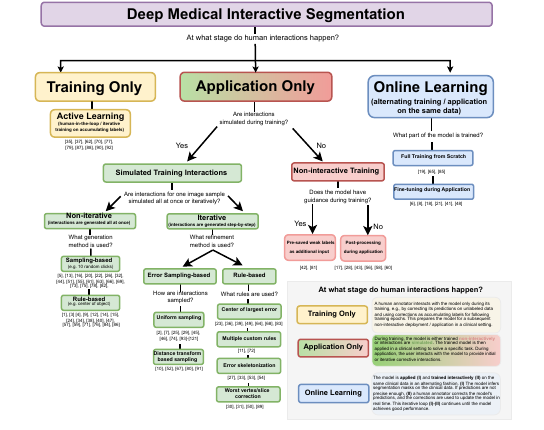}
    \caption{Our proposed taxonomy tree for all the reviewed studies. The references for studies associated with a node are listed beneath the respective node.}
    \label{fig:taxonomy}
\end{figure*}

\begin{figure*}
    \centering
    \hspace*{0.5cm}
    \includegraphics[width=1.0\linewidth]{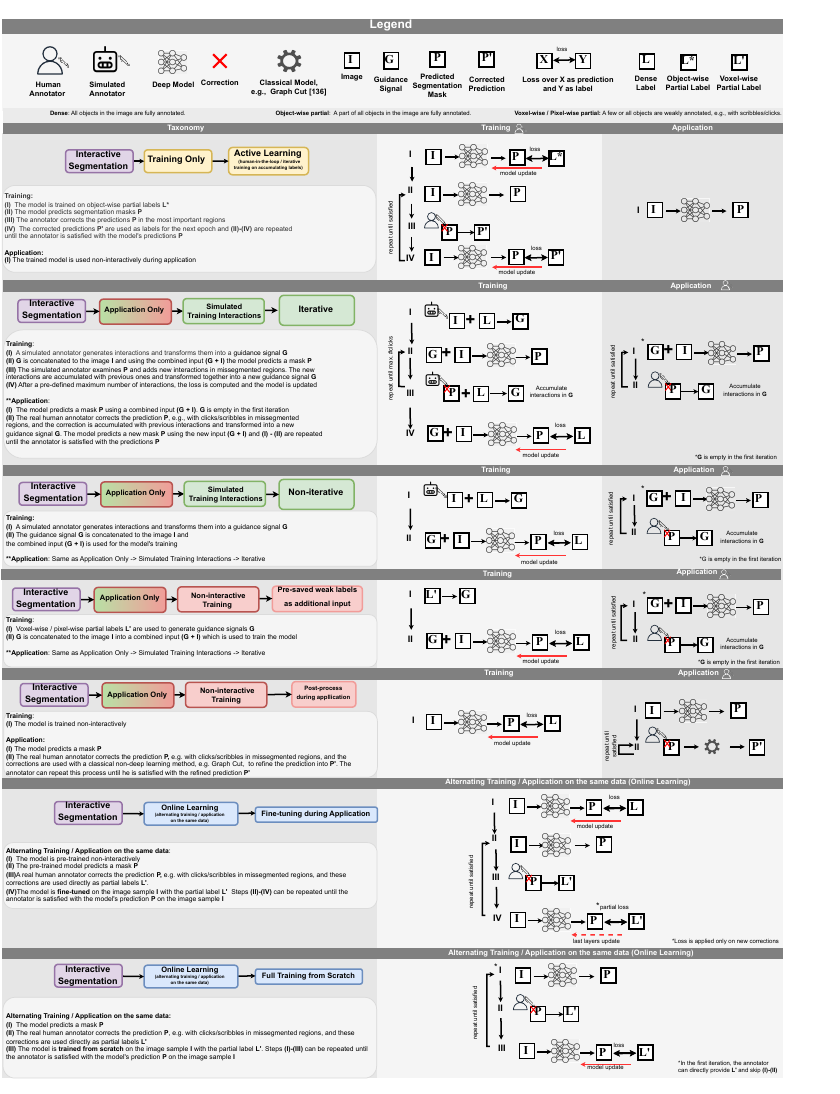}
    \caption{Taxonomy blueprints for our proposed taxonomy nodes. The human annotator is involved during training, application, or both in \textit{online learning}.}
    \label{fig:icon-table}
\end{figure*}

After retrieving the 121 publications, we analyze the foundational principles of their methodologies and categorize them based on common characteristics. This procedure yields our proposed \textit{taxonomy tree} and \textit{taxonomy blueprints} in Fig. \ref{fig:taxonomy} and \ref{fig:icon-table}, which function as navigational tools for existing medical interactive segmentation methods. These tools should help researchers categorize their approaches and steer them towards existing methods that align with their own. In this section, we provide detailed insights into the construction of both tools.

\textit{Taxonomy tree.} In our systematic review of deep medical interactive segmentation, we identified three paradigms that are determined by the stage at which human interactions occur. These paradigms form the primary categorization in our taxonomy tree in Fig. \ref{fig:taxonomy}, and a summary of each paradigm can be found in the three boxes at the bottom right. The interactions take place in two distinct stages: training and application, which are defined in Section \ref{sec:train_application}. Depending on these two stages, interactions can occur: 1) exclusively during application; 2) exclusively during training; 3) or in an alternating manner between both stages (online learning). These three paradigms constitute our proposed taxonomy and are described in detail in Sections \ref{sec:app_only}, \ref{sec:train_only}, and \ref{sec:online_learning}.

\textit{Taxonomy blueprints.} Fig. \ref{fig:icon-table} visually depicts the training and application phases of the main taxonomy nodes, using icons to represent generic concepts, such as the input image. The diagram displays the involvement of a  human annotator during either the training or application phase. The distinction between training and application phases is apparent in most paradigms, however, in the case of the \textit{online learning} paradigm, this separation is not as evident. In \textit{online learning}, the model is alternately trained and applied to the same data, with real-time feedback provided by a human annotator. The taxonomy blueprints offer a two-fold advantage: 1) they reveal detailed differences in training and application phases among nodes in the taxonomy tree; 2) and streamline the categorization of emerging methods. They serve as a visual guide to both understand the taxonomy nodes and systematically incorporate new approaches into the existing taxonomy structure.

\subsection{Training Only}
\label{sec:train_only}
The first taxonomy category encompasses methods that utilize human interactions only during training and is highlighted as yellow nodes in the taxonomy tree in Fig. \ref{fig:taxonomy}. In this paradigm, also referred to as "Active Learning", models are first trained on a small labeled fraction of the dataset ("starting budget") and are subsequently applied to the unlabeled remainder of the dataset. Based on these predictions, the most informative samples for future training are identified, annotated, and added to the training data for the next iteration. This iterative training process continues until the annotator is content with the model's predictions. Afterward, the model can be used non-interactively on the application data without the involvement of a human annotator, as seen in Fig. \ref{fig:icon-table}. Ho et al. \cite{ho2020deep} pioneer the use of active learning in the field of deep medical interactive segmentation by applying a Convolutional Neural Network (CNN) on an unlabeled osteosarcoma dataset reducing the annotation time significantly compared to standard pixelwise annotation. Menon et al. \cite{menon2021interactive} ask the annotator to highlight a query patch for the annotation of a whole-slide image (WSI). A retrieval module chooses the K-nearest patches in the image to the query patch by assessing their similarity in the feature space. The annotator then provides feedback for each patch, either as relevant or irrelevant, or provides explicit segmentation labels. This way, only the most informative patches are annotated and Menon et al. \cite{menon2021interactive} show that using their retrieval module, only 5\% of the patches need to be annotated to achieve state-of-the-art performance. Atzeni et al. \cite{atzeni2022deep} introduce a method that leverages estimated segmentation quality and labeling effort to identify regions of interest effectively. The labeling effort is based on the boundary length and its irregularity, assuming that large structures with complex boundaries are difficult to annotate. The segmentation quality is measured by the average class Dice score on already annotated regions. As a result, easier regions can be selected for annotation in the initial rounds enabling the model to acquire valuable features that prove beneficial for annotating more challenging areas at later stages. AnatomySketch \cite{zhuang2022anatomysketch} presents an open-source software platform with a graphical user interface designed for annotating and integrating deep learning segmentation models. The "Annotation-by-iterative-Deep-Learning (AIDL)" module enables annotators to proofread, correct, and incorporate segmentations into the next training iteration of a pre-trained model. Deep SED-Net \cite{liang2022deep} demonstrates how an AIDL strategy for testicular cell segmentation achieves the same results as manual annotation using squeeze-and-excitation layers \cite{hu2018squeeze} in a U-Net model \cite{ronneberger2015u}. Ma et al. \cite{ma2022rapid} utilize an igniter network trained on a small dataset to generate coarse labels on a larger dataset for a bigger model. The larger model is trained in an AIDL loop, involving a human annotator following a specific labeling protocol. The protocol prioritizes easier samples for early labeling and gradually addresses harder samples in subsequent iterations, effectively minimizing human effort while improving the larger model's final predictions on the challenging samples, similar to Atzeni et al. \cite{atzeni2022deep}. Zhuang et al. \cite{zhuang2023efficient} introduce a boundary contour correction tool as an alternative to voxel-wise corrections. Their approach demonstrates improved shape learning, faster proofreading, and more anatomically plausible results, showcasing the benefits of using a contour representation. Ho et al. \cite{ho2023deep} further accelerate the AIDL paradigm by employing a pre-trained breast segmentation model instead of using random weight initialization, resulting in reduced annotation time. Zhuang et al. \cite{zhuang2023annotation} utilize user-provided scribbles to compute an exponentialized geodesic distance map, which is then used to modulate the model's prediction. This process generates pseudo-labels for the subsequent training iteration. The pseudo-labels are implicitly more certain near the scribbles and incorporate human feedback during training. Qu et al. \cite{qu2023abdomenatlas} train three models (U-Net \cite{ronneberger2015u}, Swin-UNETR \cite{hatamizadeh2021swin}, and nnU-Net \cite{isensee2021nnu}) on a small CT dataset and use these models to annotate 8000 CT volumes. They use the inconsistency between the three models' predictions, the prediction entropy, and the overlap of model predictions to suggest volumes for refinement to the annotator. They show the three metrics correlate highly with error regions, i.e., can identify which volumes are the most informative to refine and reduce the annotation time of 8000 CTs to two work weeks \cite{qu2023abdomenatlas}.

\subsection{Application Only}
\label{sec:app_only}
The second category of our taxonomy encompasses models engaging with human annotators exclusively during the application stage, depicted as green and red nodes in Fig. \ref{fig:taxonomy}. During the training stage, these models either: 1) utilize simulated interactions generated by a simulated annotator (green nodes), often referred to as a \textit{robot user} in literature \cite{nickisch2010learning}; or 2) have no interactions (red nodes). 
During the application stage, human users interact with these models by providing initial and/or iterative corrective interactions.

\subsubsection{Simulated Training Interactions} To circumvent the need for human annotators during the training process, one approach is to simulate the annotation process using a \textit{robot user.} This robot user mimics the behavior of a human annotator and relies on ground truth labels to simulate interactions only in the correct regions. In our taxonomy, we differentiate between iterative and non-iterative simulation.

\hspace*{2em}1a) \textit{Non-iterative simulation:} In the non-iterative simulation, interactions are generated all at once in a single iteration and then transformed into a guidance signal, which is combined with the image. During training, there are no correction loops, whereas during the application stage,  human annotators can iteratively correct the model's predictions, as illustrated in Fig. \ref{fig:icon-table}. Non-iterative methods can be further subdivided into the two subcategories \textit{rule-based} and \textit{sampling-based}, depending on whether the interactions are generated through deterministic rules (e.g., the center of the largest connected component in the mask) or by randomly sampling the ground-truth mask, respectively.

\textit{Non-iterative rule-based} methods employ deterministic rules to simulate interactions. For instance, Sun et al. \cite{sun2017point} simulate a click by using the center of the prostate. They utilize the Canny edge detector \cite{canny1986computational} to generate horizontal and vertical location prior maps. These maps assign decreasing intensity values to voxels that are farther from the central click and have more edges crossed (an example is given in the Appendixes). Khan et al. \cite{khan2019extreme} utilize the extreme points of the object (topmost, leftmost, rightmost, and bottom-most points) as four clicks and generate a confidence map based on the Chebyschev and Mahalanobis distances to the center of the object. DeepCut \cite{rajchl2016deepcut} extends the ground-truth bounding box to generate foreground and background voxels, which serve as inputs for an interactive CNN. After that, dense CRFs \cite{krahenbuhl2011efficient} refine the CNN's predictions, and the refined predictions from the dense CRFs are utilized again as the foreground and background voxel seeds for the interactive CNN. Can et. al. \cite{can2018learning} also utilize dense CRFs to refine CNN predictions for prostate and cardiac structures segmentation. iW-Net \cite{aresta2019iw} simulates two clicks by selecting the two farthest points in the ground truth mask. These two points are used to compute an attraction field, inspired by the punctual electric charges of opposite values (an example is given in the Appendixes). Roth et al. \cite{roth2019weakly, roth2021going} utilize 3D Gaussian heatmaps centered at the extreme points which are subsequently expanded and used as a guidance signal for a CNN-based model. Raju et al. \cite{raju2020user} tackle the domain gap between simulated and ground-truth extreme points by training a model to predict these points on unseen data and utilize them as a guidance signal. Girum et al. \cite{girum2020fast} also employ extreme points as input to their prior-knowledge network. Their network generates a spatial attention map that is then multiplied with the image and fed into a downstream segmentation model. MIDeepSeg \cite{luo2021mideepseg} leverages extreme points to simulate clicks and slightly shifts them towards the inner side of the boundary to obtain interior margin points. These points are then used to compute an exponentialized geodesic distance map as a guidance signal. Zhang et al. \cite{zhang2021interactive} extract image patches along rays extending from the object's center to its outer boundaries, which are then used to train a Convolutional Recurrent Neural Network (ConvRNN) model. In the application stage, a single click at the object's center is sufficient, as the ConvRNN model sequentially segments the neighboring patches around the click. Trimpl et al. \cite{trimpl2021interactive} simulate the complete annotation of the central axial slice in a CT image and propagate it to the remaining slices in the volume. Their model leverages the central slice, its annotation, and a target slice as joint inputs to learn the segmentation of the target slice. The entire volume is segmented by iteratively choosing each slice from the CT as a new target slice. Jahanifar et al. \cite{jahanifar2021robust} skeletonize the ground-truth mask to produce simulated scribbles. i3Deep \cite{gotkowski2022i3deep} randomly samples multiple slices from the ground-truth labels for each image sample. The full image labels of the selected slices are then appended to the input of a refinement model during the training stage. Galisot et al. \cite{galisot2022visual} train segmentation models on a variety of brain structures using cropped regions from the entire brain as inputs. Moreover, they develop a model to learn spatial relationships between structures, automatically positioning bounding boxes during inference, and allowing annotators to adjust the bounding boxes as needed. Chen et al. \cite{chen2022balancing} generate 2D Gaussian heatmaps around each extreme point and compute an Euclidean distance transform using the intersection point of the two extreme axes as a seed point to produce a second guidance signal. Bruzadin et al. \cite{bruzadin2023learning} propagate foreground seeds from a source slice to neighboring slices by considering the strong edges in the image and avoiding sampling seeds near those edges in the adjacent slices. 
Shahin et al. \cite{shahin2023sparse} identify the slice with the highest error and use the ground-truth boundary of that slice as a simulated scribble.

\textit{Non-iterative sampling-based} methods sample the ground-truth labels to simulate interactions. DeepIGeoS \cite{wang2018deepigeos} samples a fixed amount of voxels from connected components that are over a certain size threshold and uses them as seeds for computing a geodesic distance transform. UGIR \cite{wang2020uncertainty}, Bi et al. \cite{bi2022hyper}, DeepIGeoSv2 \cite{lei2019deepigeos}, WDTISeg \cite{li2021wdtiseg}, and Hallitschke et al. \cite{hallitschke2023multimodal} follow the same sampling strategy as DeepIGeoS \cite{wang2018deepigeos}. UGIR \cite{wang2020uncertainty} additionally estimates the segmentation uncertainty by calculating the prediction variance within a group convolution layer. Bi et al. \cite{bi2022hyper} integrate the guidance signal at multiple stages in their skin lesion segmentation model. DeepIGeoSv2 \cite{lei2019deepigeos} expands upon the two-stage DeepIGeoS \cite{wang2018deepigeos} model to handle multiple organs and introduces an uncertainty-aware loss function that assigns an exponential penalty based on the model's certainty of an error. WDTISeg \cite{li2021wdtiseg} combines geodesic and Euclidean distance maps through a linear combination, allowing the incorporation of both appearance and location information, respectively. Hallitschke et al. \cite{hallitschke2023multimodal} extend DeepIGeoS \cite{wang2018deepigeos} to multimodal PET/CT data and investigate different ways to present the annotation interface to the user when displaying multimodal data for annotation. Cerrone et al. \cite{cerrone2019end} segment neuron cells from serial section electron microscopy images by randomly sampling a click from each neuron in the image while enforcing a minimum distance to any boundary. Wang et al. \cite{wang2019interactive} perturb the vertices of the ground-truth polygon of the object's boundary by applying randomly sampled offsets and directions. NuClick \cite{koohbanani2020nuclick} randomly samples a point inside each nucleus so that it is at least two pixels away from the object boundaries. Tang et al. \cite{tang2020one} perform dilation on the ground-truth masks of liver and lung lesions as well as lymph nodes, and then randomly sample five pixels from the dilated mask. Jiang et al. \cite{jiang2021residual} adopt a two-stage network approach and randomly sample clicks from the segmentation error of the first-stage coarse network, encoding them as Gaussian heatmaps. Daulatabad et al. \cite{daulatabad2021integrating} sample multiple clicks in the proximity of the centroid of the ground-truth mask of the thyroid nodule. Shi et al. \cite{shi2022hybrid} partition the ground-truth mask into sections based on the Euclidean distance to the object's boundary. Subsequently, one random pixel is sampled from each section. This technique mitigates the issue of cluttered samples that are in close proximity to each other in the guidance map. Shahedi et al. \cite{shahedi2022effect} and Ju et al. \cite{ju2022all} uniformly sample clicks from the target organ. Both methods experiment with different numbers of sampled clicks in their ablations studies. Pirabaharan et al. \cite{pirabaharan2022improving, pirabaharan2022interactive} uniformly sample the ground-truth mask to generate foreground and background clicks. These clicks are encoded using Gaussian heatmaps, where the radius of each heatmap is proportional to the area of the corresponding ground-truth mask. The heatmaps exhibit smaller radii for smaller objects, allowing them to align more accurately with their boundaries.

\hspace*{2em}1b) \textit{Iterative simulation:} Iterative simulation methods mimic the iterative nature of interaction segmentation during application where the human annotator repeatedly corrects the prediction on the model in a typical human-in-the-loop scenario. This loop is simulated by either sampling interactions from the missegmented regions or defining deterministic rules to choose each next interaction, e.g., choosing the center of the largest erroneous region. 

\textit{Iterative error sampling-based} methods sample interactions for the next iteration from missegmented regions. We distinguish between uniform and distance transform-based sampling in our taxonomy, depending on how interactions are sampled. 

\textit{Iterative uniform sampling} approaches sample new interactions with an equal probability of landing in any of the missegmented pixels/voxels. UI-Net \cite{amrehn2017ui} identifies the missegmented regions of hepatic lesions and samples a random number of pixels for each interaction. For the first interaction, they initialize the foreground and background scribbles by performing multiple dilation and erosion operations on the lesion's boundary, respectively. InterCNN \cite{bredell2018iterative} uniformly samples multiple clicks from the error, places a $9\times9$ window around each click, and adds all foreground pixels in the window to the sampled clicks, regardless of whether they were missegmented. Hu et al. \cite{hu2020error} use a stratification approach, randomly sampling a click from each of the three largest missegmented connected components. Li et al. \cite{li2021interactive} randomly sample clicks from the intersection area between the object's boundary and the error regions. They adopt a reinforcement learning approach, where the agent is rewarded based on the cross-entropy improvement. They also propose a confidence estimation network that guides the human annotator during application by suggesting click locations based on the segmentation confidence. Deng et al. \cite{deng20213d} adopt a sampling strategy where a fixed number of under- and oversegmented voxels are selected at each iteration. Their loss function exclusively focuses on the $9 \times 9 \times 9$ neighborhood surrounding each missegmented voxel to avoid contaminating the already well-segmented regions. Mikhailov et al. \cite{mikhailov2022deep} randomly sample clicks from missegmented regions and store the sampled clicks from all iterations in an ordered memory bank. The memory bank preserves the sequence of interactions instead of combining all the clicks into a single guidance signal, ensuring that the sequential information is retained.

Recently, Meta AI released the code for their Segment Anything Model (SAM) \cite{kirillov2023segment}. Due to its remarkable performance and zero-shot capabilities on natural images, many methods have adapted SAM for medical images. In this review, we only consider methods that use SAM's interactive prompts. All the reviewed medical SAM methods fall into the category of iterative uniform sampling simulation in our taxonomy, which uses SAM's original pre-training described in the training algorithm in [\citenum{kirillov2023segment}, p.17]. Here, we summarize these methods.

Mazurowski et al. \cite{mazurowski2023segment} extensively evaluate the zero-shot performance of SAM on 33 datasets and observe significant performance variations across different tasks, ranging from 0.11 to 0.86 Intersection over Union. Their findings highlight that bounding box prompts consistently yield superior results and that SAM perfroms better on larger objects. Interestingly, the study demonstrates that iterative corrections do not lead to substantial improvements, with the best performance achieved in the first three clicks for most tasks. Deng et al. \cite{deng2023segment} similarly conclude that SAM excels in segmenting larger objects but faces challenges when dealing with multiple small objects, even with an abundance of prompts. The study highlights that SAM is unsuitable for gigapixel WSI data. SAM vs. BET \cite{mohapatra2023sam} exhibits SAM's superior performance over the gold standard Brain Extraction Tool (BET) \cite{jenkinson2005bet2} in brain extraction from Magnetic Resonance Imaging (MRI) images but it does not compare it to more recent skull stripping models \cite{he2023accuracy}. Putz et al. \cite{putz2023segment} demonstrate SAM's effective generalization in glioma brain tumor segmentation, except for small tumors under $300$ mm$^3$, where its performance shows some deterioration. Hu et al. \cite{hu2023sam} examine SAM's effectiveness in liver tumor segmentation, but the study concludes that there exists a considerable performance gap compared to even a simple U-Net model \cite{ronneberger2015u}. SAM-Adapter \cite{chen2023sam} incorporates task-specific knowledge by injecting task-specific embeddings into SAM's image encoder which leads to a significant improvement in SAM's performance for polyp segmentation compared to using it directly without any modifications. Ophthalmology SAM \cite{qiu2023learnable} fine-tunes SAM with an additional prompt adapter on fundus images and improves SAM significantly on three ophthalmology tasks.

He et al. \cite{he2023accuracy} evaluate SAM on 12 public medical datasets spanning ten organs and six imaging modalities. Their findings reveal that SAM is consistently outperformed by a simple U-Net \cite{ronneberger2015u} across all 12 datasets, and its performance is notably influenced by the size of the target object. Moreover, SAM achieves notably higher results on 2D modalities (dermoscopy, colonoscopy, X-Ray, ultrasound) when compared to 3D modalities such as MRI and CT. Shi et al. \cite{shi2023generalist} confirm that SAM is easily outperformed by a simple U-Net model \cite{ronneberger2015u} in fundus, CT, MRI, and Optical Coherence Tomography (OCT) data. Nonetheless, they showcase that through in-domain fine-tuning, SAM reaches the performance level of specialized U-Net models in retinal vessel segmentation. GazeSAM \cite{wang2023gazesam} employs eye gazing to estimate the annotator's point of focus, encodes these positions as clicks, and utilizes SAM for segmentation. SkinSAM \cite{hu2023skinsam} fine-tunes SAM on dermoscopy images by using simulated bounding box prompts, resulting in satisfying performance on skin lesion segmentation. Wang et al. \cite{wang2023sam} utilize SAM for surgical instrument segmentation and find that bounding box-based prompting outperforms click prompts by a significant margin. However, SAM's performance remains unsatisfactory in challenging scenarios, such as dealing with overlapping instruments and blood. Cheng et al. \cite{cheng2023sam} assess SAM without fine-tuning on 12 medical datasets, demonstrating that bounding boxes yield significantly better results compared to clicks. Additionally, the study finds that incorporating perturbations into the bounding boxes leads to a deterioration in performance. 
Mattjie et al. \cite{mattjie2023exploring} investigate SAM across six datasets and affirm that employing the ground-truth bounding box without perturbations yields the best results consistently across all datasets and across various transformer backbones of SAM. Polyp-SAM \cite{li2023polyp} fine-tunes SAM on five colonoscopy datasets using bounding box prompts. They find that fine-tuning only the decoder instead of the whole model and using a smaller transformer backbone achieves the best performance. BreastSAM \cite{hu2023breastsam} also concludes that using a smaller transformer backbone leads to better results for breast cancer segmentation in ultrasound images. IAMSAM \cite{lee2023iamsam} implements an annotation interface for microscopy images where the segmentation masks are utilized for downstream tasks, such as cell type prediction and spatial transcriptomics. Shen et al. \cite{shen2023temporally} extend SAM by incorporating temporal prompts, where a Reinforcement Learning (RL) agent suggests the appropriate type of prompt, such as a bounding box or click. The study demonstrates that the learned RL suggestions outperform choosing only one of the types of interaction. Ning et al. \cite{ning2023potential} apply SAM to ultrasound videos and unveil its potential for segmenting various structures, showcasing minimal deviations between video frames when sufficient prompts are provided. Zhang et al. \cite{zhang2023segment} conduct experiments on multiple anatomical structures and conclude that SAM performs most effectively on large structures like the liver and brain. However, its performance deteriorates when applied to smaller and ambiguous targets such as the parotid and cochlea. MedLSAM \cite{lei2023medlsam} uses extreme points that implicitly define a bounding box prompt for SAM and reduce the annotation burden. SAM-U \cite{deng2023sam} generates multiple bounding box prompts for a single input image. By computing the entropy of the predictions obtained by using each bounding box in a separate forward pass, SAM-U estimates the aleatoric uncertainty. This uncertainty metric highlights challenging regions in the image that necessitate additional guidance from the annotator. 3DSAM-adapter \cite{gong20233dsam} is the first to adapt SAM to 3D images and 3D prompts. This adaptation is achieved by freezing the pre-trained weights and extending each of SAM's components to 3D. The patch embedding is extended using a 3D depth-wise convolution. The 3D position encoding is produced by summing the embedding from SAM's original 2D lookup table with the embedding from a new depth lookup table. The attention block extends from $[B, H \times W, c]$ to $[B, H \times W \times D, c]$ queries, and the bottleneck replaces all 2D convolutions with 3D. Huang and Yang et al. \cite{huang2023segment} curate 52 public datasets to assess SAM's "Segment Everything" mode and different click and bounding box prompts. Their evaluation reveals that SAM's performance varies considerably across the datasets, even for the same structure in different modalities. Additionally, the study concludes that using bounding boxes leads to higher and more consistent performance compared to using clicks, and employing the "Segment Everything" mode yields the least favorable results.

In contrast to the predominantly negative findings in most other works that integrate SAM for medical images, MedSAM \cite{ma2023segment} achieves a remarkable performance on 14 unseen datasets, covering 50 target classes and seven imaging modalities, and even surpasses specialized nnU-Net \cite{isensee2021nnu} models on each of the target classes. This impressive outcome is the result of the careful curation of 84 existing public medical datasets for pre-training, leading to 1\,090\,486 medical image-mask pairs, and fine-tuning SAM on this large-scale medical dataset. The diversity of this dataset, spanning 15 imaging modalities, bolsters MedSAM's strong generalization abilities and reveals the significant potential of using SAM for medical interactive segmentation. Furthermore, MedSAM \cite{ma2023segment} concludes that bounding box prompts perform the best, and they convert 3D images into 2D slices for training and evaluation. Medical Sam Adapter (MSA) \cite{wu2023medical} implements an adapter that extends to depth attention to account for the dimensionality reduction from 3D to 2D images in SAM's training. They pre-train SAM on multiple large-scale medical datasets and show superior performance to MedSAM \cite{ma2023segment} but only when using clicks instead of bounding boxes. 
PromptUNet \cite{wu2023promptunet} proposes to use U-Net \cite{ronneberger2015u} as a backbone for all encoders and decoders in SAM instead of a Vision Transformer (ViT) \cite{dosovitskiy2020image}. PromptUNet \cite{wu2023promptunet} shows that this architecture change leads to a significant improvement, outperforming both click-based MedSAM \cite{ma2023segment} and MSA \cite{wu2023medical} on five datasets. 
DeSAM \cite{gao2023desam} disentangles the prompt from the image to avoid the influence of poor prompts and shows considerable improvement to MedSAM \cite{ma2023segment}, even with bounding box prompts.

\textit{Iterative distance transform-based sampling} approaches apply a distance transform over the missegmented regions, generating a distance map that serves as a sampling distribution for new interactions. As a result, these approaches prioritize sampling new interactions primarily in the central regions of the connected components of the errors. Sakinis et al. \cite{sakinis2019interactive} introduced an approach that applies the Chamfer distance transform to errors and use the distance map as a sampling distribution for foreground and background clicks in each iteration. Bai et al. \cite{bai2021progressive} apply the Euclidean distance transform to the over- and undersegmented regions to create the background and foreground sampling distributions respectively. They exponentialize and normalize the distance maps to convert them to a pseudo-probability map. DeepEdit \cite{diaz2022deepedit} adopts the same strategy as Sakinis et al. \cite{sakinis2019interactive}. However, they also experiment with different proportions of interaction-free iterations where the model receives no clicks, i.e., an empty guidance signal. Bai et al \cite{bai2022proof} apply the Euclidean distance transform to the error regions, followed by a Softmax normalization to produce a pseudo-probability map to sample new clicks. Guiding the Guidance (GtG) \cite{marinov2023guiding} extends DeepEdit \cite{diaz2022deepedit} and proposes a dynamic Gaussian heatmap that has a larger radius in homogeneous regions and a smaller radius near edges. They compare four existing guidance signals and propose five evaluation metrics to facilitate a more systematic comparison of interactive models.

\textit{Iterative rule-based} approaches utilize a deterministic rule to generate an interaction at each iteration. We differentiate between four types of rules: 1) center of largest error; 2) error skeletonization; 3) worst vertex/slice correction; and 4) multiple custom rules.

\textit{Center of largest error.} Several methods use the center of the largest error region as the next click with the assumption that it is the most intuitive choice. IterMRL\cite{liao2020iteratively} and BS-IRIS \cite{ma2020boundary} employ multi-agent reinforcement learning where each voxel is an agent with a cross-entropy improvement reward. During each iteration, IterMRL \cite{liao2020iteratively} selects the center of the largest error region along with the centers of the other $N-1$ largest connected components. Feng et al. \cite{feng2021interactive} combine few-shot learning with interactive segmentation by using only a small set of annotated slices and assigning clicks only to them during training. New clicks at each iteration are placed in the center of the largest connected error component of these slices. DINs \cite{zhang2021dins} position new clicks in the center of the largest error region and verify if it belongs to the ground-truth class. For concave error regions where the center is not part of the error, an additional step is performed. The error region is skeletonized, and the nearest point in the skeleton to the selected point is used as the final click location. This ensures accurate click placement even in concave error regions. iSegFormer \cite{liu2022isegformer} introduces a transformer-based interactive model for knee cartilage segmentation and samples a click at the center of the under- and oversegmentated regions. Liu et al. \cite{liu2022transforming} also introduce a transformer-based architecture that expands beyond binary segmentation to handle multiple classes. To address missegmentation, their approach involves placing new clicks at the centers of the largest missegmented regions for each class separately. Liu et al. \cite{liu2023exploring} incorporate cycle consistency to preserve the quality of the initial segmentation in the refinement steps. They simulate the initial click in the largest anatomical structure and refine only the worst segmented organ with central clicks in the following steps. 

\textit{Error skeletonization} simulates iterative scribbles as an alternative to iterative clicks. Similar to the central error clicks, error skeletonization generates scribbles that are positioned in the central regions of the object. A visual example of error skeletonization is given in the Appendixes. Kitrungrotsakul et al. \cite{kitrungrotsakul2020interactive} simulate foreground and background scribbles through skeletonization of under- and oversegmented errors, respectively. They utilize an initial non-interactive model followed by a second-stage model that incorporates the generated scribbles as additional input. Jinbo et al. \cite{jinbo2020development} build upon the work of Kitrungrotsakul et al. \cite{kitrungrotsakul2020interactive} by introducing an enhanced annotation interface. This interface allows users to draw scribbles while simultaneously visualizing the preliminary segmentation result. DeepScribble \cite{cho2021deepscribble} computes  Euclidean distance maps for the false positive and false negative regions. The distance maps are thresholded and skeletonized to simulate scribble-like annotations. Attention-RefNet \cite{kitrungrotsakul2021attention} also adopts skeletonized errors to emulate scribbles. The guidance signal is computed by subtracting the geodesic distance maps of the foreground and background so that it assigns positive values to the foreground and negative values to the background.

\textit{Worst vertex/slice correction.} Another deterministic rule for enhancing segmentation involves identifying and selecting the worst vertex or slice at each iteration and incorporating its ground-truth value as a guidance signal. Tian et al. \cite{tian2020graph} leverage this approach by predicting the boundary polygon vertices and simulating the user dragging the worst vertex towards its correct position. They utilize a Graph Convolutional Network (GCN) to propagate this information to the remaining vertices in the polygon and update the segmentation contour accordingly.
Tian et al. \cite{tian2021interactive} extend this approach by incorporating a local correction. They update only the $2\times K$ neighboring vertices of the corrected vertex. By confining the update to the local neighborhood, the other parts of the contour remain preserved, preventing changes to already well-segmented regions. Chao et al. \cite{chao2020interactive} adopt a similar approach, but instead of focusing on individual vertices, they target the worst segmented 2D slice. The corrections of the worst slice serve as a supervisory signal to update the bottleneck feature of the segmentation model, after which the model is applied again to the whole volume. Wei et al. \cite{wei2023towards} also simulate a slice correction and use the ground-truth label of the slice with the largest tumor area as a guidance signal. They compare this strategy to random slice corrections to show that their approach selects more informative slices.

\textit{Multiple custom rules.} The last taxonomy tree node we assign to the green partition are methods that apply multiple custom rules that are specific to the application. Zhou et al. \cite{zhou2019interactive} apply different types of interactions and simulate them during training. They simulate a central click by eroding the largest connected component multiple times and then select the center from the remaining eroded pixels. They also simulate scribbles by selecting the worst segmented 2D slice and connecting the two farthest points in the largest error component. Lin et al. \cite{lin2022multi} simulate a boundary scribble by dilating the object's boundary and simulate iterative clicks by placing a central click in the largest error.

\subsubsection{Non-interactive Training} Several interactive methods opt to exclude interactions during the training stage and, instead, adopt a standard non-interactive training approach. These methods are marked in red in Fig. \ref{fig:taxonomy}. Based on their approach, these methods either post-processing the model prediction during the application or incorporate additional weak labels during training.

\hspace*{2em}\textit{2a) Post-processing during application:} Some methods adopt non-interactive training and integrate post-processing techniques to combine model predictions with user interactions during application. For instance, Zheng et al. \cite{zheng2019deep} employ shadow set theory \cite{pedrycz1998shadowed} to extract all ground-truth masks from the training dataset, which are fully aligned with the human clicks on an unseen image during application. The extracted masks are averaged and fused with the model's prediction for the unseen image, while the mask variance serves as an estimate of uncertainty. In IRIS \cite{pepe2020iris}, patches centered around user clicks are extracted from the full volume and fed into a pre-trained model. The final prediction is obtained by stitching together the resulting predictions from all patches in a post-processing step. Williams et al. \cite{williams2021interactive} employ B-spline active surfaces \cite{barbosa2011b} to calculate an active contour along the boundary of a CNN prediction. The user can modify the contour by dragging its control points, and the active contour is updated by minimizing the Yezzi energy \cite{yezzi2002fully} with respect to the pixel intensities. PiPo-Net \cite{fang2021pipo} utilizes a two-stage model comprising a U-Net \cite{ronneberger2015u} to generate a pixelwise segmentation mask and a post-processing Long Short-Term Memory (LSTM) model \cite{hochreiter1997long} to produce a vertex polygon over the mask's boundary. The polygon vertices are predicted in consecutive steps. During application, the user can correct vertices, and the LSTM model updates all consecutive vertices as a post-processing step. Manh et al. \cite{manh2021interactive} use a U-Net \cite{ronneberger2015u} for Z-line segmentation, followed by post-processing with a Binary Partition Tree (BPT) \cite{salembier2000binary}. Users mark superpixels \cite{achanta2012slic} with foreground and background clicks, and the BPT resolves conflicts with the U-Net predictions based on the Euclidean distance to the labeled superpixels. 
Sun et al. \cite{sun2022efficient} adopt a two-stage approach for boundary prediction. First, a CNN predicts the initial contour, and subsequently, a GCN is trained as a post-processing step to predict the offset of the predicted boundary vertices from the ground-truth boundary.

\hspace*{2em}\textit{2b) Pre-saved weak labels as additional input.} Some models utilize additional weak labels during training. However, instead of using the weak labels as supervisory signals as done in weakly supervised learning \cite{zhou2018brief}, the weak labels are transformed into guidance signals. Zhou et al. \cite{zhou2021quality, zhou2023volumetric} utilize weak labels in the form of scribbles on a single source slice and propagate the label information from the source slice to the rest of the volume using a memory-readout operation from a memory-encoder network. 

\subsection{Online Learning}
\label{sec:online_learning}
The third category in the taxonomy tree encompasses methods that undergo real-time training or fine-tuning directly on the data they are finally applied to. Methods in this paradigm produce on-the-fly predictions and allow annotators to make immediate corrections with minimal or no latency between corrections, model updates, and new predictions. In our taxonomy, we differentiate between full training and fine-tuning based on the number of updated model parameters.

\subsubsection{Full Training from Scratch} Some models do not use any pre-training and are trained entirely on the data on which they are finally applied. These models use the user interactions as the only labels, update their parameters in real-time, and predict again so that the user can correct them again until they are satisfied with the prediction's quality. Längkvist et al. \cite{langkvist2019interactive} present a real-time annotation tool designed for CT scans to segment pulmonary fibrosis by training CNNs from scratch using only human interactions as labels. Their study investigates the trade-off between accuracy and efficiency by examining the performance of small and large models in online learning. ECONet \cite{asad2022econet} utilizes a small CNN model with only one convolution layer. After the user draws a scribble, ECONet uses fixed-size patches centered around each voxel in the scribble for model updates by utilizing the provided scribbles directly as the ground-truth mask. After each weight update, the model is applied to the whole volume using sliding window inference for Coronavirus Disease 2019 (COVID-19) lung lesion segmentation. Asad et al. \cite{asad2023adaptive} extend ECONet \cite{asad2022econet} by proposing an adaptive loss that propagates the influence of scribbles to neighboring regions with similar features. They also prune voxels with predictions under a certain confidence threshold to discard uncertain samples during weight updates.

\subsubsection{Fine-tuning during Application} The second option for online learning models is to utilize a pre-trained model and only fine-tune it on the application data using human interactions. BIFSeg \cite{wang2018interactive} utilizes a user-provided bounding box to crop the target and generates an initial segmentation using a pre-trained model. The user then corrects the initial predictions using scribbles, and the model undergoes fine-tuning with a scribble-based weighted loss function. Dhara et al. \cite{dhara2019segmentation} expand the fine-tuning step into an iterative loop, where the annotator corrects model predictions with scribble-based GraphCut \cite{boykov2001interactive}. These corrections are then used to update the model in real-time. Chao et al. \cite{chao2019radiotherapy} propagate user corrections on a slice to neighboring slices by updating the model with a loss function based on the distance to the corrected slice. Boers et al. \cite{boers2020interactive} employ a loss function that assigns higher weights to voxels from user-provided scribbles that are missegmented, while other voxels are weighted based on their distance to the scribbles. 
Sambaturu et al. \cite{sambaturu2021efficient} showcase an efficient model-agnostic scheme for fine-tuning using user-based scribbles. They dilate the scribbles with region growing and introduce an L2 regularization term to update the model's weights. The L2 term ensures that the updated weights are not drastically different from the initial ones, promoting stability in the model's predictions during the fine-tuning process.

\section{Review Findings}
\label{sec:findings}
In this section, we present our findings on the prevalent trends observed during the review of the 121 reviewed papers. We delve into the implications of these trends and the potential factors contributing to them. Through this analysis, we aim to provide a comprehensive depiction of the current landscape within the medical interactive segmentation domain.

\begin{figure*}[!t]
\hspace*{2.0cm}
    \begin{adjustbox}{minipage=\linewidth,scale=0.85}

\centering
\subfigure{\includegraphics[width=0.458\linewidth]{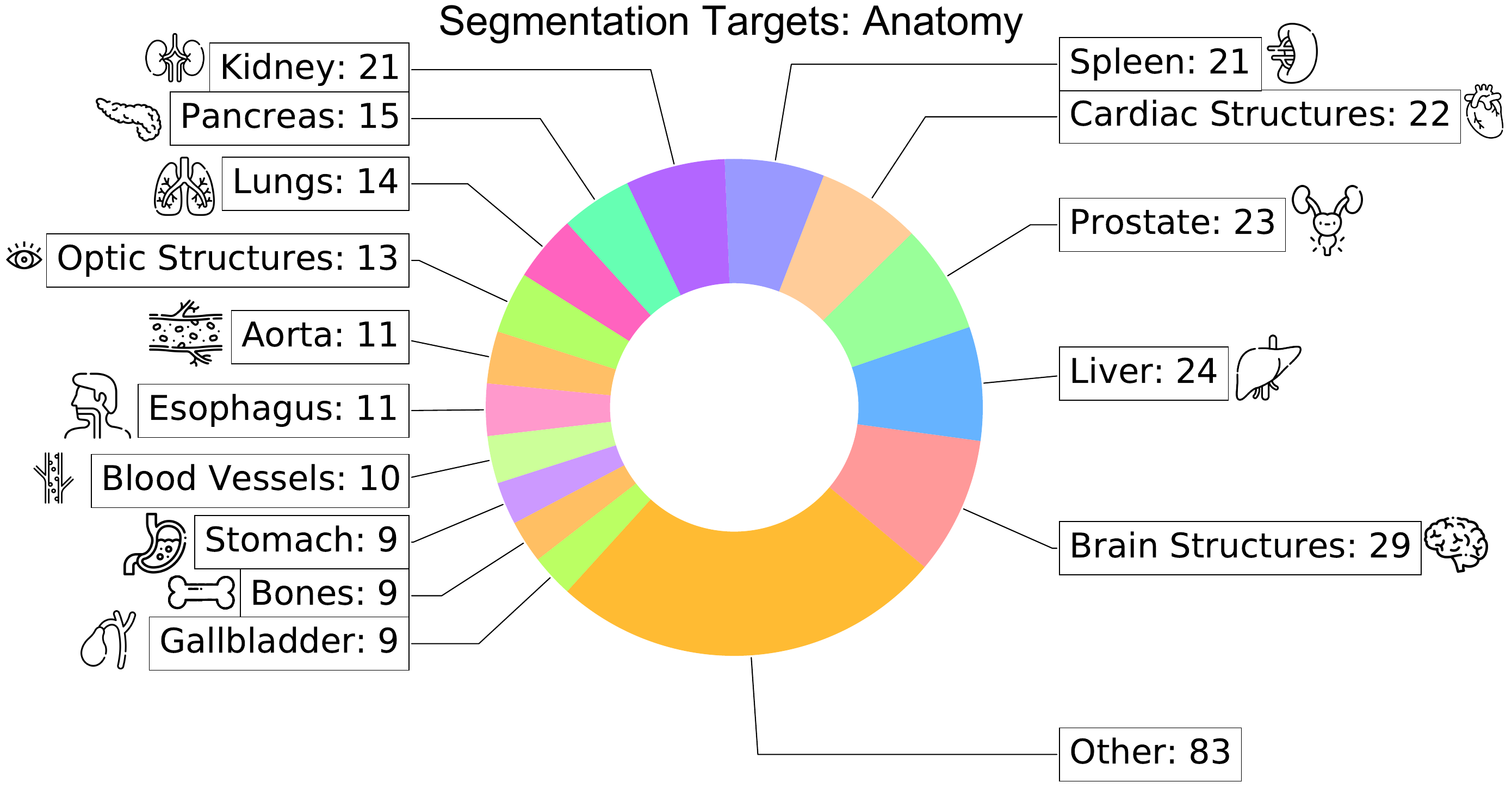}
  }
  \subfigure{\includegraphics[width=0.478\linewidth]{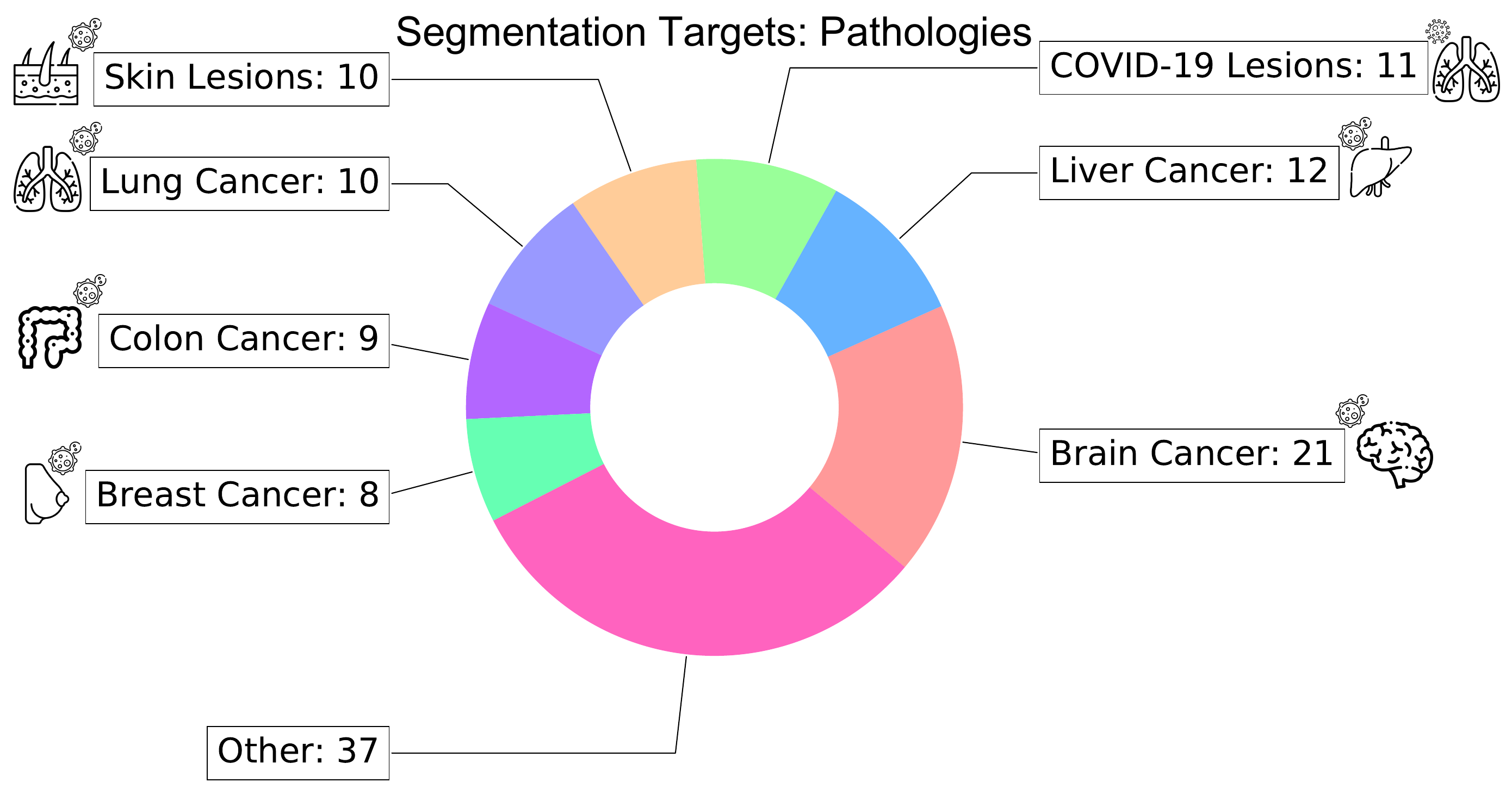}
  }

\centering
\hspace*{-0.53cm}
  \subfigure{\includegraphics[width=0.448\linewidth]{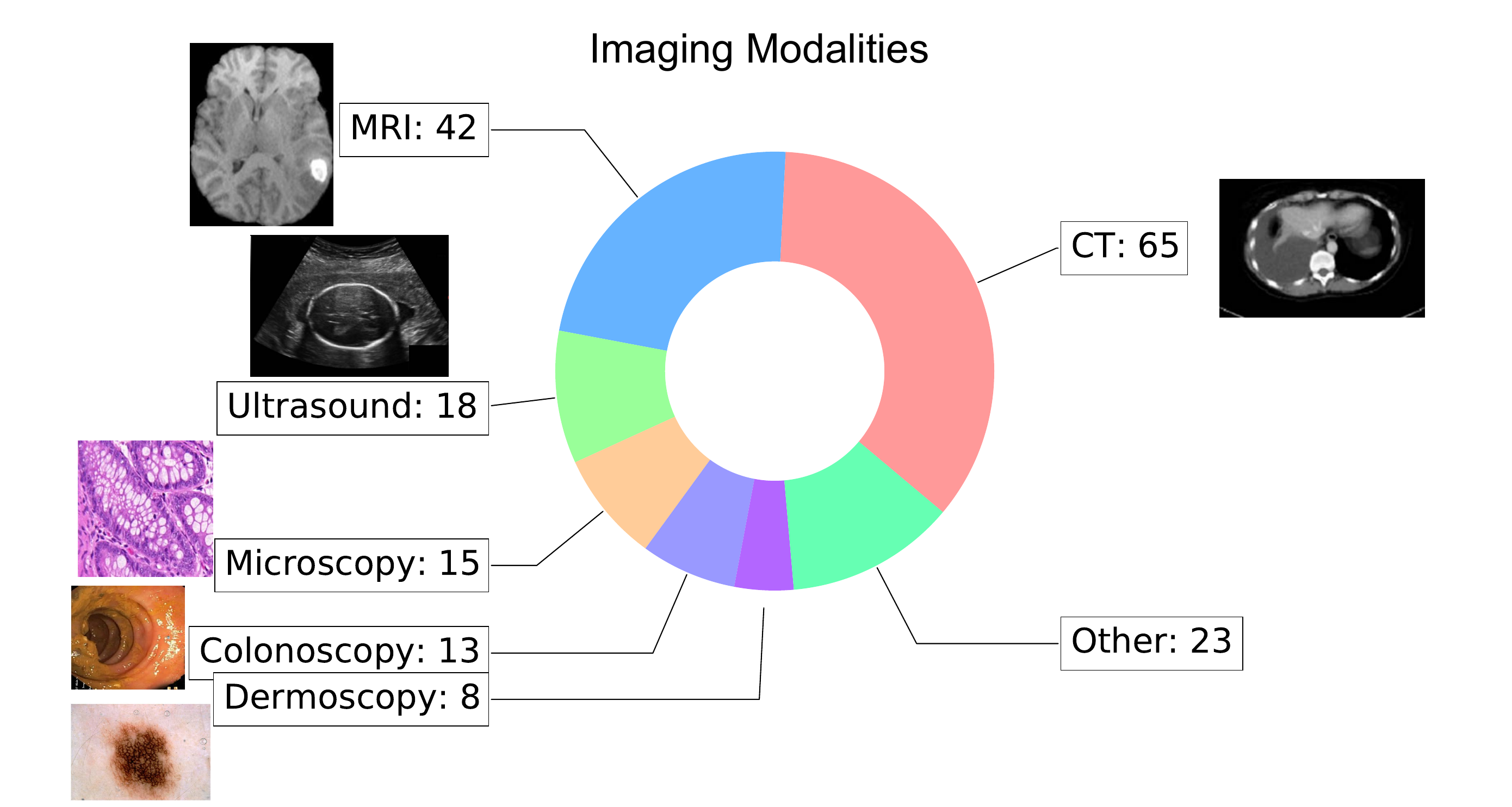}
  } \hspace*{0.1cm}
  \subfigure{\includegraphics[width=0.468\linewidth]{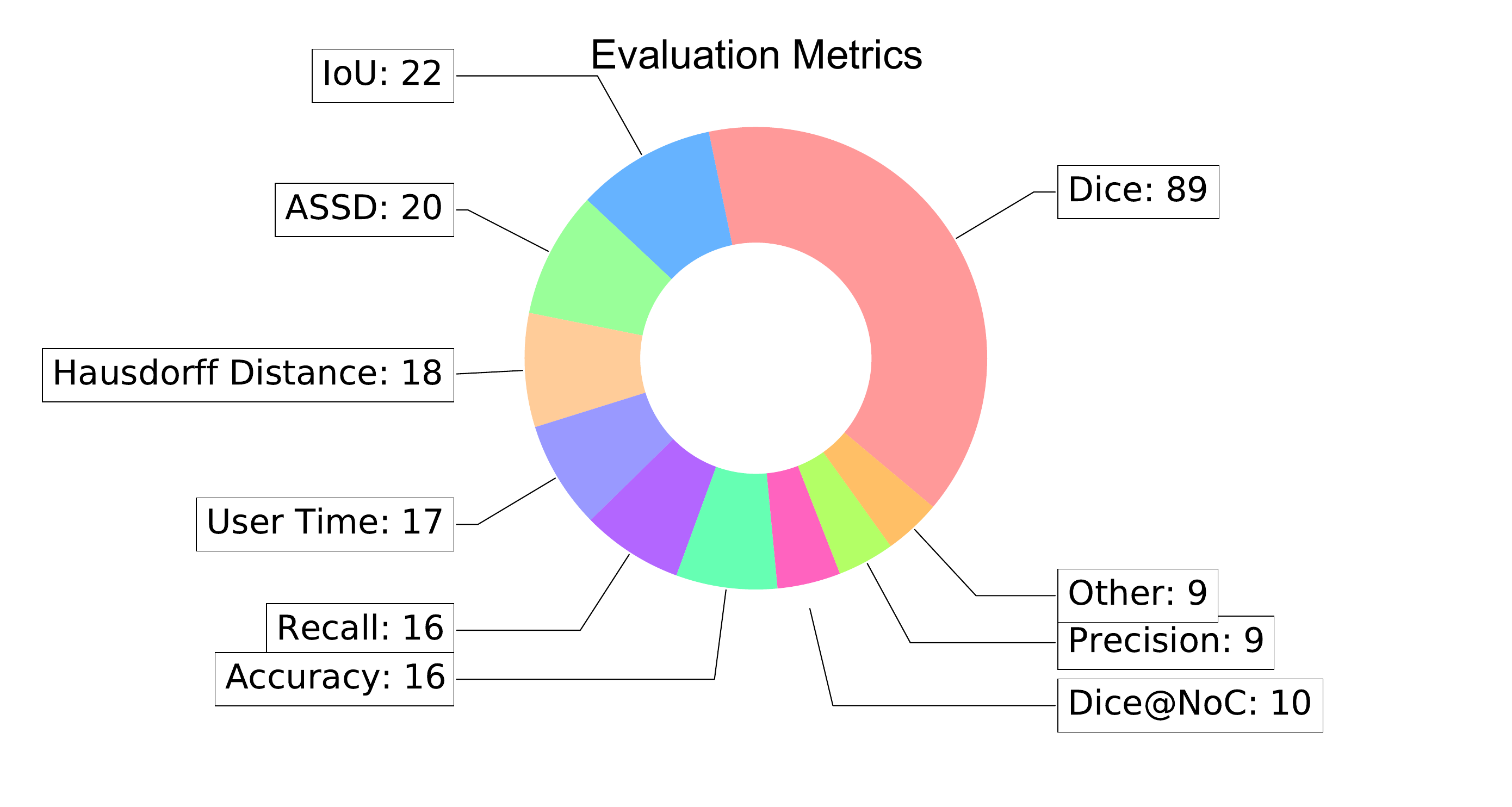}
  }

\end{adjustbox}

\caption{Distribution of segmentation targets in the anatomy (top left) and in the pathology (top right), imaging modalities (bottom left), and evaluation metrics (bottom right) among all reviewed papers. The numbers represent the number of papers in that category. The icons on the top row are designed by \href{https://www.flaticon.com}{Flaticon.com}.}
\label{fig:pie_plots_2}
\end{figure*}
% Icons made by Freepik from www.flaticon.com.

\subsection{Segmentation Targets, Imaging Modalities, and Evaluation Metrics}
\subsubsection{Segmentation Targets} We distinguish segmentation targets in two primary categories: 1) anatomical structures and cells; and 2) pathologies. The categorization depends on whether a method's primary focus is on specific anatomical structures, distinct pathologies, or both (noted in n=7 of all 121 studies). The number of methods per specific anatomy or pathology is depicted in Fig. \ref{fig:pie_plots_2}. Prominent anatomical regions encompass the brain, prostate, and cardiac structures as well as abdominal organs featuring the liver, spleen, kidney, pancreas, stomach, and gallbladder. Thoracic organs are less prominent, including lungs, aorta, esophagus, and cardiac structures, and whole-body structures like bones and blood vessels. Further notable regions of interest which are combined in the "Other"-category encompass lymph nodes, the Z-line, spine, cartilage, and skin. Furthermore, techniques using microscopy and OCT data are predominantly geared towards cell segmentation, targeting blood cells, testicular cells, neurons, or cell nuclei.

Pathological targets exhibit notably less diversity compared to their anatomical counterparts. The prevalent pathologies tend to concentrate primarily within the brain (n=21) and liver regions (n=12), largely owing to the prominence of datasets like: 1) BraTS \cite{menze2014multimodal} for brain cancer; 2) as well as MSD \cite{antonelli2022medical} and LiTS \cite{bilic2023liver} for liver cancer. Beyond brain and liver cancer, a few specific targets emerge as representative of certain imaging modalities. Notably, COVID-19 lung lesions stand out in X-Rays, while skin lesions take precedence in dermoscopy. Colon cancer and polyps serve as typical examples in colonoscopy imaging. Other relevant pathologies encompass lung, breast, kidney, and thyroid cancer. The "Other"-category consists of less frequently encountered targets such as head and neck cancer, cervical, pancreatic, prostatic, and esophageal cancer, hematomas, and foot ulcers.

\subsubsection{Imaging Modalities}  Radiological modalities, particularly CT (n=65) and MRI (n=42), dominate the imaging modalities and are featured in the most reviewed methods. This prevalence can be attributed to the existence of popular public datasets from segmentation challenge competitions like MSD \cite{antonelli2022medical} and BraTS \cite{menze2014multimodal}. These challenges frequently release their training data publicly, incentivizing the adoption of these imaging modalities in many approaches. Subsequent to CT and MRI, ultrasound is the choice for n=18 out of 121 approaches, frequently applied in cardiac imaging, mammography, or fetal ultrasound. Microscopy finds application in n=15 out of 121 reviewed methods, predominantly in pathology for tumor or cancer cell identification. Colonoscopy stands as an imaging modality exclusively dedicated to polyp and/or colon cancer segmentation. Dermoscopy, on the other hand, specializes in skin lesion segmentation. Less frequently encountered imaging modalities in interactive models encompass OCT, X-Ray, fundus imaging, and PET/CT. For a comprehensive listing of segmentation targets and imaging modalities utilized by each reviewed method, refer to Tables \ref{tab:paper_taxonomy} and \ref{tab:paper_taxonomy_sam}.

\subsubsection{Evaluation Metrics} %Fig. \ref{fig:pie_plots_2} visually represents the evaluation metrics employed by the surveyed approaches. The Dice similarity coefficient emerges as the predominant choice, widely embraced for its well-established reputation in measuring the final segmentation performance \cite{maier2022metrics}. Following closely behind the Dice score, interactive segmentation methods frequently report outcomes using metrics like Intersection over Union (IoU), Average Symmetric Surface Distance (ASSD), the Hausdorff Distance (HD), Recall, Voxel/Pixel Accuracy, and Precision, providing a quantitative assessment of performance after all interactions.
An adequate selection of evaluation metrics is crucial for meaningful assessment of segmentation methods, and thus, for trustworthy deployment in practice as well as scientific progress of the field. A large-scale investigation recently found that current medical image segmentation is subject to a substantial extent of pitfalls related to evaluation metrics \cite{reinke2023understanding}. The study reveals various shortcomings of the popular Dice Similarity Coefficient (DSC). At the same time, a follow-up study termed “Metrics Reloaded” provides a standardized framework for avoiding these pitfalls and selecting adequate metrics for a given problem \cite{maier2022metrics}. One major finding was that performance should always be assessed by multiple metrics to account for failure modes such as of the DSC.  Fig. \ref{fig:pie_plots_2} depicts the evaluation metrics employed by the reviewed studies. As expected, the most-used metric is DSC (n=89). However, oftentimes the DSC is the only reported metric for segmentation performance (n=29). Another common problem is the reporting of redundant metrics, such as reporting both DSC and Intersection over Union (n=19). Remarkably, despite its widespread use as a complementary metric to DSC in non-interactive segmentation and its endorsement by "Metrics Reloaded" for various settings, the Normalized Surface Distance \cite{nikolov2021clinically} appears in only 2 out of 121 studies reviewed.

The incorporation of user-centered metrics is crucial for devising user-friendly and intuitive methods, particularly in the context of human-in-the-loop approaches. However, there is a noticeable scarcity of user-centric metrics in the reviewed studies. Some studies report the User Time (n=17), quantifying the active annotator's labeling time in seconds, or the Dice@NoC (n=10), measuring the Dice score at a predefined Number of Clicks (NoC). Furthermore, the "Other"-category includes usability metrics like NASA-TLX \cite{hart1988development} and the System Usability Scale \cite{brooke1996sus}, although these are seldom utilized.

\begin{figure*}[t!]
    \centering
   \begin{tikzpicture}
    \node[anchor=south west, inner sep=0] (image) at (0,0){\includegraphics[width=0.8\textwidth]{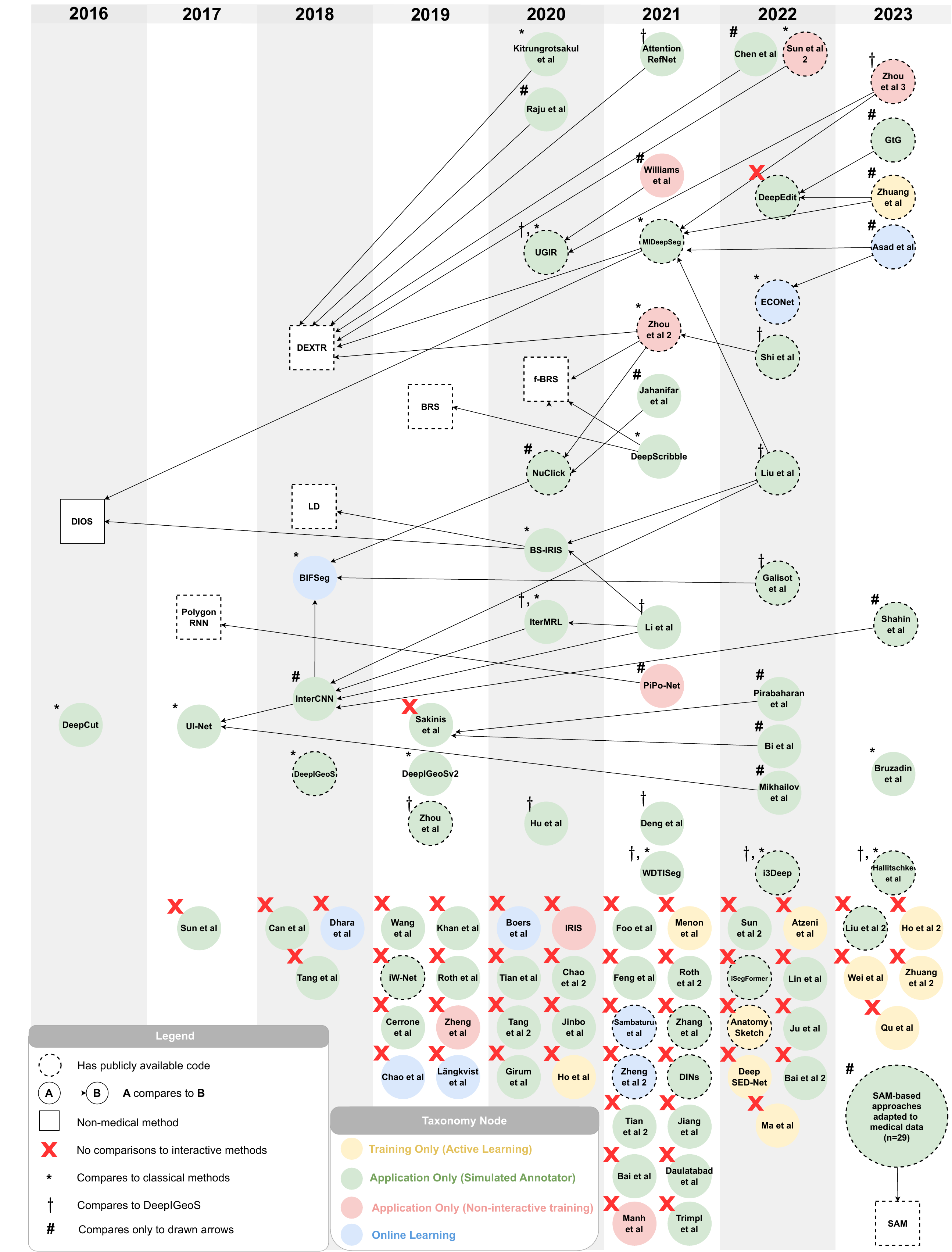}};
\begin{scope}[x={(image.south east)},y={(image.north west)}]

    % 2016
      \node at (0.11,0.56) {\tiny\cite{xu2016deep}};

      \node at (0.11,0.41) {\tiny\cite{rajchl2016deepcut}};
    % 2017
      \node at (0.23,0.482) {\tiny\cite{castrejon2017annotating}};

      \node at (0.23,0.402) {\tiny\cite{amrehn2017ui}};

      \node at (0.235,0.245) {\tiny\cite{sun2017point}};
    % 2018
      \node at (0.35,0.6975) {\tiny\cite{maninis2018deep}};

      \node at (0.35,0.6175) {\tiny\cite{li2018interactive}};

      \node at (0.35,0.52) {\tiny\cite{wang2018interactive}};

      \node at (0.35,0.4225) {\tiny\cite{bredell2018iterative}};

      \node at (0.35,0.36) {\tiny\cite{wang2018deepigeos}};

      \node at (0.325,0.245) {\tiny\cite{can2018learning}};

      \node at (0.38,0.245) {\tiny\cite{dhara2019segmentation}};

      \node at (0.355,0.2075) {\tiny\cite{tang2018semi}};

    % 2019

      \node at (0.475,0.65) {\tiny\cite{jang2019interactive}};

      \node at (0.475,0.405) {\tiny\cite{sakinis2019interactive}};

      \node at (0.475,0.365) {\tiny\cite{lei2019deepigeos}};

      \node at (0.475,0.3225) {\tiny\cite{zhou2019interactive}};
    
      \node at (0.4475,0.2475) {\tiny\cite{wang2019interactive}};

      \node at (0.505,0.2475) {\tiny\cite{khan2019extreme}};

      \node at (0.4475,0.2075){\tiny\cite{aresta2019iw}};

      \node at (0.505,0.2075) {\tiny\cite{roth2019weakly}};

      \node at (0.4475,0.1695){\tiny\cite{cerrone2019end}};

      \node at (0.505,0.1695) {\tiny\cite{zheng2019deep}};

      \node at (0.4475,0.1295){\tiny\cite{chao2019radiotherapy}};

      \node at (0.505,0.1295) {\tiny\cite{langkvist2019interactive}};

    % 2020

      \node at (0.5975,0.94) {\tiny\cite{kitrungrotsakul2020interactive}};

      \node at (0.5975,0.925) {\tiny\cite{raju2020user}};

      \node at (0.5975,0.82) {\tiny\cite{wang2020uncertainty}};

      \node at (0.5975,0.72) {\tiny\cite{sofiiuk2020f}};

      \node at (0.5975,0.6025) {\tiny\cite{koohbanani2020nuclick}};

      \node at (0.5975,0.5775) {\tiny\cite{ma2020boundary}};

      \node at (0.5975,0.49) {\tiny\cite{liao2020iteratively}};

      \node at (0.5975,0.325) {\tiny\cite{hu2020error}};

      \node at (0.5725,0.2475) {\tiny\cite{boers2020interactive}};

      \node at (0.63,0.2475) {\tiny\cite{pepe2020iris}};

      \node at (0.5725,0.2075){\tiny\cite{tian2020graph}};

      \node at (0.63,0.2075) {\tiny\cite{chao2020interactive}};

      \node at (0.5725,0.1695){\tiny\cite{tang2020one}};

      \node at (0.63,0.1695) {\tiny\cite{jinbo2020development}};

      \node at (0.5725,0.1295){\tiny\cite{girum2020fast}};

      \node at (0.63,0.1295) {\tiny\cite{ho2020deep}};

    % 2021
      \node at (0.72,0.94) {\tiny\cite{kitrungrotsakul2021attention}};
    
      \node at (0.72,0.88) {\tiny\cite{williams2021interactive}};
    
      \node at (0.7275,0.7925) {\tiny\cite{luo2021mideepseg}};
    
      \node at (0.72,0.72) {\tiny\cite{zhou2021quality}};

      \node at (0.72,0.67) {\tiny\cite{jahanifar2021robust}};

      \node at (0.72,0.62) {\tiny\cite{cho2021deepscribble}};

      \node at (0.72,0.485) {\tiny\cite{li2021interactive}};

      \node at (0.72,0.44) {\tiny\cite{fang2021pipo}};

      \node at (0.72,0.325) {\tiny\cite{hu2020error}};

      \node at (0.72,0.285) {\tiny\cite{li2021wdtiseg}};
    
      \node at (0.6925,0.2475) {\tiny\cite{foo2021interactive}};

      \node at (0.75,0.2475) {\tiny\cite{menon2021interactive}};

      \node at (0.6925,0.2075) {\tiny\cite{feng2021interactive}};

      \node at (0.75,0.2075) {\tiny\cite{roth2021going}};

      \node at (0.6925,0.1675) {\tiny\cite{sambaturu2021efficient}};

      \node at (0.75,0.1675) {\tiny\cite{zhang2021interactive}};

      \node at (0.6925,0.1295) {\tiny\cite{zheng2021continual}};

      \node at (0.75,0.1295) {\tiny\cite{zhang2021dins}};

      \node at (0.6925,0.0875) {\tiny\cite{tian2021interactive}};

      \node at (0.75,0.0875) {\tiny\cite{jiang2021residual}};

      \node at (0.6925,0.0495) {\tiny\cite{bai2021progressive}};

      \node at (0.75,0.0495) {\tiny\cite{daulatabad2021integrating}};

      \node at (0.6925,0.0075) {\tiny\cite{manh2021interactive}};

      \node at (0.75,0.0075) {\tiny\cite{trimpl2021interactive}};

    % 2022
      \node at (0.8125,0.94) {\tiny\cite{chen2022balancing}};
    
      \node at (0.8725,0.94) {\tiny\cite{sun2022efficient}};
    
      \node at (0.8425,0.86) {\tiny\cite{diaz2022deepedit}};
    
      \node at (0.8425,0.7425) {\tiny\cite{asad2022econet}};
    
      \node at (0.8425,0.695) {\tiny\cite{shi2022hybrid}};
    
      \node at (0.8425,0.605) {\tiny\cite{liu2022transforming}};
    
      \node at (0.8425,0.515) {\tiny\cite{galisot2022visual}};
    
      \node at (0.8525,0.425) {\tiny\cite{pirabaharan2022interactive, pirabaharan2022improving}};
    
      \node at (0.8425,0.39) {\tiny\cite{bi2022hyper}};
    
      \node at (0.8425,0.35) {\tiny\cite{mikhailov2022deep}};
    
      \node at (0.8425,0.285) {\tiny\cite{gotkowski2022i3deep}};
    
      \node at (0.815,0.2475) {\tiny\cite{sun2022efficient}};

      \node at (0.8725,0.2475) {\tiny\cite{atzeni2022deep}};

      \node at (0.815,0.2075) {\tiny\cite{liu2022isegformer}};

      \node at (0.8725,0.2075) {\tiny\cite{lin2022multi}};

      \node at (0.815,0.1675) {\tiny\cite{zhuang2022anatomysketch}};

      \node at (0.8725,0.1675) {\tiny\cite{ju2022all}};

      \node at (0.815,0.1275) {\tiny\cite{liang2022deep}};

      \node at (0.8725,0.1275) {\tiny\cite{bai2022proof}};

      \node at (0.84,0.0875) {\tiny\cite{ma2022rapid}};

    % 2023
      \node at (0.9625,0.9175) {\tiny\cite{zhou2023volumetric}};
    
      \node at (0.9625,0.87) {\tiny\cite{marinov2023guiding}};
    
      \node at (0.9625,0.825) {\tiny\cite{zhuang2023efficient}};
    
      \node at (0.9625,0.785) {\tiny\cite{asad2023adaptive}};
    
      \node at (0.9625,0.48) {\tiny\cite{shahin2023sparse}};
    
      \node at (0.9625,0.365) {\tiny\cite{bruzadin2023learning}};
    
      \node at (0.9625,0.285) {\tiny\cite{hallitschke2023multimodal}};
    
      \node at (0.935,0.2475) {\tiny\cite{liu2023exploring}};

      \node at (0.9925,0.2475) {\tiny\cite{ho2023deep}};

      \node at (0.935,0.2075) {\tiny\cite{wei2023towards}};

      \node at (0.9925,0.2075) {\tiny\cite{zhuang2023annotation}};

      \node at (0.96,0.1625) {\tiny\cite{qu2023abdomenatlas}};

      \node at (0.9975,0.0725) {\tiny\cite{mazurowski2023segment} --\cite{roy2023sam}};

      \node at (0.9775,0.0075) {\tiny\cite{kirillov2023segment}};

    \end{scope}
    \end{tikzpicture}
 
    \caption{Comparison graph of all the reviewed methods. Nodes are ordered temporally from left to right. Classical methods denote non-deep learning-based interactive methods before 2016. The star ($*$) and the dagger ($\dagger$) are introduced to reduce the visual load in the figure caused by too many arrows.}
    \label{fig:comparison_graph}
\end{figure*}
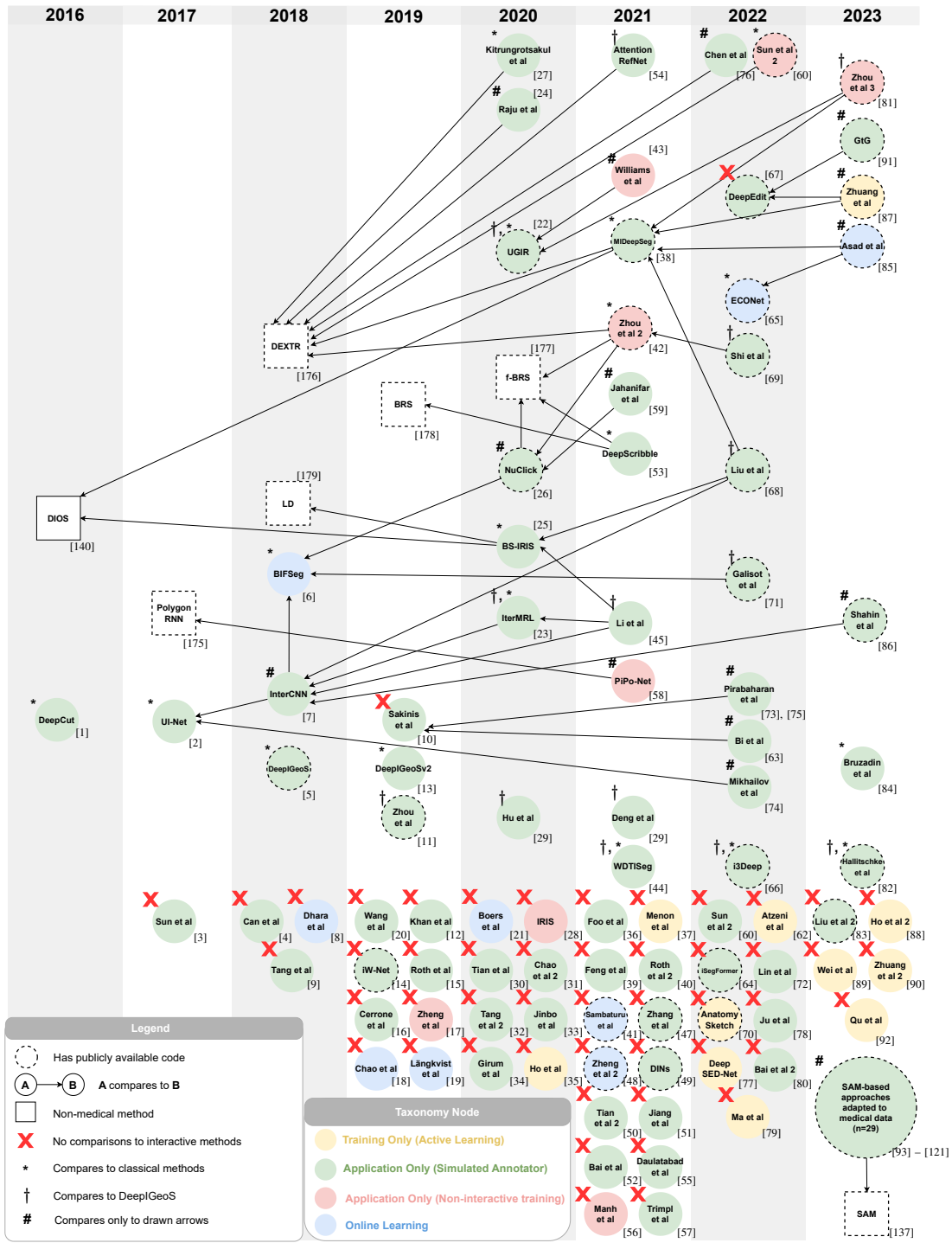

\subsection{Emergence of Foundation Models}
In early 2023, the Segment Anything Model (SAM) \cite{kirillov2023segment} emerged, introducing an approach that involves large-scale training on over 1 billion segmentation masks. Although SAM's initial training dataset (SA-1B) primarily comprises 2D natural images, several works have showcased its adaptability to medical data, spanning both 2D (such as dermoscopy and fundus) and 3D imaging modalities (including CT, MRI, and PET/CT). This versatility is achieved through targeted fine-tuning on medical data \cite{ma2023segment}. In the case of 3D images, it commonly involves using 2D axial slices \cite{ma2023segment} or integration of specialized 2D-to-3D adapters into the model \cite{gong20233dsam}. 

SAM has shown a good generalization on multiple imaging modalities and tasks, especially on 2D modalities \cite{ma2023segment} utilizing its bounding box prompting capability. This light-weight adaptability has caused an unprecedented acceleration in the field of deep interactive medical image segmentation as evidenced by 29 proposed medical SAM-adaptations in only a few months at the time of writing. Thus, SAM has demonstrated the potential of utilizing foundation models for medical interactive segmentation. Further, due to its generalization and zero-shot capabilities, it seems to foster a trend towards evaluating methods on a larger number of tasks as some SAM-based approaches are evaluated on over 30 public medical datasets \cite{ma2023segment, huang2023segment}. 

\subsection{Reproducibility and Availability}
In recent years, the field of interactive medical segmentation has witnessed a surge in the emergence of new approaches. There is a promising shift towards enhanced reproducibility, with an increasing number of research papers releasing their code, often accompanied by detailed instructions for replicating results, and in some instances, providing pre-trained model weights. This openness and transparency in sharing resources are further bolstered by the presence of open-source projects like MONAI Label \cite{diaz2022monai}, AnatomySketch \cite{zhuang2022anatomysketch}, RIL-Contour \cite{philbrick2019ril}, BioMedisa \cite{losel2020introducing}, MITK \cite{wolf2005medical}, and PyMIC \cite{wang2023pymic} which greatly facilitate the development and deployment of interactive deep medical models. Non-deep learning projects such as ilastik \cite{sommer2011ilastik}, ITK-Snap \cite{yushkevich2016itk}, and Li et al. \cite{li2022efficient}, have also contributed to the open source development of interactive models and are widely used in the community. Additionally, this positive trajectory benefits from a growing reliance on openly available challenge datasets sourced from platforms such as Kaggle (\href{www.kaggle.com}{www.kaggle.com}), Grand Challenge (\href{www.grand-challenge.org}{www.grand-challenge.org}), and Synapse (\href{www.synapse.org}{www.synapse.org}), promoting collaborative research and advancing the state of the art in the domain. All these tendencies are illustrated in Fig. \ref{fig:num_pub}. In our Appendixes, we provide links to code repositories of all reviewed studies with publicly available code as well as links to all 185 public datasets used by the reviewed methods, streamlining access for future researchers.

\subsection{Comparison Graph}
Finally, we investigate the field’s practice of comparing proposed methods against relevant baselines. Since the scientific merit of a proposed method is measured as the gain over existing solutions, a comprehensive and up-to-date set of baseline methods is crucial for scientific progress in the field. Fig. \ref{fig:comparison_graph} gives an overview over the comparison practices in deep interactive segmentation of medical images. 

The most remarkable observation is the fact that a large fraction (n=46) of the 121 reviewed studies \textit{do not compare against any prior work}. Another portion compares exclusively against “classical methods” (n=6), i.e. non-deep learning-based methods proposed before 2016, or exclusively against DeepIGeoS \cite{wang2018deepigeos} (n=3). Additionally, a large portion of studies (n=37) compare only to interactive methods which are not trained on medical data, such as DIOS \cite{xu2016deep}, Polygon-RNN \cite{castrejon2017annotating}, DEXTR \cite{maninis2018deep}, Latent Diversity (LD) \cite{li2018interactive}, BRS \cite{jang2019interactive}, f-BRS \cite{sofiiuk2020f}, and SAM \cite{kirillov2023segment}.  Despite their shared characteristics, even methods from the same node of the presented taxonomy tree (see circle color in Fig. \ref{fig:comparison_graph}) are most often not compared against each other. Finally, the described acceleration of the field caused by the introduction of SAM seems to compromise the rigor of evaluation given that none of these approaches compares to methods other than the original non-medical SAM. This overall concerning status of a severe lack of cross-comparison in the field comes as a surprise given the positive trends towards reproducibility shown in Fig. \ref{fig:num_pub}.

\section{Discussion and Future Directions}
Based on the key trends we have identified in Section \ref{sec:findings}, we now derive and discuss the major challenges and opportunities for the field of deep interactive medical image segmentation. The discussion aims to provide a succinct summary of the field's current trajectory while simultaneously identifying pivotal areas where course corrections are necessary.

\subsection{Positive Trends}
\subsubsection{Momentum in Research and Adaptation} The increasing number of publications each year reflects significant momentum and rapid advancements in the field. Additionally, the fast adoption of new paradigms, such as SAM \cite{kirillov2023segment}, exemplifies the field's dynamic and responsive nature to emerging concepts and technologies.
\subsubsection{Enhanced Reproducibility and Open-Source Engagement} There has been a notable surge in the use of open-source methods and public datasets. This trend not only facilitates more accessible development of customized models but also encourages the sharing of these models within the community. The proliferation of open-source frameworks specifically designed for interactive segmentation, like MONAI Label \cite{diaz2022monai} and AnatomySketch \cite{zhuang2022anatomysketch}, further underlines this commitment to reproducibility and collaborative growth.

\subsection{Challenges and Opportunities}

Our review highlights a pivotal challenge in the field: a discernible deficiency in scientific rigor in the evaluation of methods. These shortcomings are evident in various aspects, which we discuss in the following alongside opportunities to address them.
\subsubsection{Missing Baselines and Scattered Comparisons} The absence of consistent baselines and scattered comparisons across studies is a major issue. Frequently, new methods are not compared with previous work, possibly due to a lack of awareness of other methods or no established evaluation protocols for interactive segmentation.

\textit{Opportunities:} First, we hope that our taxonomy tree functions as a navigational tool, aiding researchers in categorizing their approaches and guiding them towards relevant existing methods. Second, the emergence of generalizing models like SAM \cite{kirillov2023segment} is a promising trend towards foundational baselines that allow for out-of-the-box comparisons under a uniform protocol. This approach can shift the field towards more structured and systematic improvements, similar to the effects of nnU-Net \cite{isensee2021nnu} in the realm of non-interactive medical segmentation, which, due to its out-of-the-box functionality, serves as a strong and standardized baseline in the field \cite{eisenmann2023winner}.
\subsubsection{No Standardized Benchmarking Datasets} The lack of established benchmarking protocols across datasets and tasks in interactive medical image segmentation is a significant barrier. This gap impedes the objective evaluation and comparison of interactive models, which results in an inconsistent literature landscape with no definite state of the art.

\textit{Opportunities:} The domain of non-medical interactive segmentation, particularly with natural images, has addressed this issue by leveraging extensively validated benchmark datasets like GrabCut \cite{rother2004grabcut}, DAVIS \cite{perazzi2016benchmark}, Pascal VOC \cite{everingham2010pascal}, SBD \cite{hariharan2011semantic}, and Berkeley \cite{mcguinness2010comparative}. Moreover, these datasets are coupled with well-defined evaluation protocols and metrics, streamlining fair and systematic comparisons with previous research. A potential remedy for the fragmented nature of comparisons within the medical interactive segmentation field entails the establishment of a curated selection of the most exemplary datasets tailored to specific tasks and imaging modalities, complete with well-defined evaluation protocols. Such an approach would furnish researchers with a systematic framework for assessing their methodologies and documenting enhancements over prior methods. 
\subsubsection{Lack of Adequate and Standardized Evaluation Metrics} In the current landscape of deep interactive medical image segmentation, there are two significant challenges related to metric selection. The first prevalent issue is the over-reliance on a single metric for evaluating segmentation performance. As pointed out in \cite{reinke2023understanding, maier2022metrics}, this approach is too narrow and often fails to adequately capture the complexity and nuances of segmentation accuracy. Second, there is a conspicuous absence of user-centric metrics in evaluations. These metrics are essential to understand how effectively an interactive segmentation tool meets the practical needs and scenarios of its users, especially in the medical imaging context.

\textit{Opportunities:} By adopting the comprehensive guidelines of "Metrics Reloaded" for metric selection, researchers can ensure a more holistic evaluation of segmentation methods. This would involve using a diverse set of metrics that together provide a more complete picture of a method's performance. In addition to technical metrics, emphasizing user-centric metrics in evaluations is crucial. This focus will shed light on the usability and practical effectiveness of interactive segmentation methods from the perspective of end-users, which is particularly important in clinical applications.

\section{Conclusion}

In conclusion, our systematic review and the accompanying taxonomy tree stand as a pivotal resource for both researchers and practitioners within the field of deep interactive medical image segmentation. For researchers, this work simplifies the task of locating pertinent related studies, thereby enhancing the quality and relevance of their methodological proposals and evaluations. Practitioners, meanwhile, are empowered to swiftly identify and select methods that are optimally suited to their unique problem scenarios. Additionally, our review has not only identified key trends within the field but also thoroughly discussed the related challenges and opportunities for the future. Most importantly, we have pinpointed a concerning lack of scientific rigor in the evaluation of methods. This critical insight underlines the need for more standardized and systematic benchmarking practices in the field. Overall, we believe this work represents an important step towards implementing such standardized approaches, thereby fostering the development of more reliable, efficient, and effective solutions in deep interactive medical image segmentation.

\bibliographystyle{ieeetr} % <===========================================

\bibliography{egbib}

%\vfill
%\clearpage
\end{document}